\documentclass[twocolumn, twocolappendix]{aastex631}

\usepackage{chngcntr}
\usepackage{enumitem}
\usepackage{amsmath}
\usepackage{txfonts}
\usepackage{xcolor}
\usepackage{graphicx}
\usepackage{gensymb}
\usepackage{booktabs}
\usepackage{bm}
\usepackage{lipsum}
\usepackage{fnpct}
\usepackage{algorithm}
\usepackage{algpseudocode}
\usepackage{pgffor}
\usepackage{soul}

\AtBeginDocument{\mathcode`v=\varv}

\providecommand{\sorthelp}[1]{} % used with the Planck_bib.bib file

\newcommand{\NHI}{\ifmmode N_{{\mathrm{H}} \, \mathrm{I}} \else $N_{{\mathrm{H}} \, \mathrm{I}}$\fi} 
\newcommand{\HI}{\ifmmode \mathrm{\ion{H}{1}} \else \ion{H}{1} \fi}
\newcommand{\CII}{\ifmmode \mathrm{\ion{C}{2}} \else \ion{C}{2} \fi}
\newcommand{\ROHSA}{{\tt ROHSA}}

\newcommand\ab{\bm{a}}

\newcommand\rb{\bm{r}}

\newcommand\mub{\bm{\mu}}
\newcommand\sigmab{\bm{\sigma}}

\def\GHz{\ifmmode $\,GHz$\else \,GHz\fi}
\def\MJysr{\ifmmode \,$MJy\,sr\mo$\else \,MJy\,sr\mo\fi}
\def\microns{\ifmmode \,\mu$m$\else \,$\mu$m\fi}

\def\kms{\ifmmode $\,km\,s$^{-1}\else \,km\,s$^{-1}$\fi}

\defcitealias{marchal_2019}{M19}
\defcitealias{blagrave_dhigls:_2017}{B17}

\newcommand{\am}[1]{%
  {\color{blue} #1} %
}

\received{February 13, 2023}
\revised{August 11, 2023}
\accepted{November 21, 2023}
\submitjournal{ApJ}

\shorttitle{Mapping a lower limit on the mass fraction of the CNM using Fourier transformed \HI\ 21\,cm emission line spectra}
\shortauthors{Marchal et al.}

\graphicspath{{./}{figures/}}

\begin{document}

\title{Mapping a lower limit on the mass fraction of the cold neutral medium using Fourier transformed \HI\ 21\,cm emission line spectra: Application to the DRAO Deep Field from DHIGLS and the HI4PI survey}

\correspondingauthor{Antoine Marchal}
\email{antoine.marchal@anu.edu.au}

\author[0000-0002-5501-232X]{Antoine Marchal}
\affiliation{Canadian Institute for Theoretical Astrophysics, University of Toronto, 60 St. George Street, Toronto, ON M5S 3H8, Canada}
\affiliation{Research School of Astronomy \& Astrophysics, Australian National University, Canberra ACT 2610 Australia}

\author[0000-0002-5236-3896]{Peter G. Martin}
\affiliation{Canadian Institute for Theoretical Astrophysics, University of Toronto, 60 St. George Street, Toronto, ON M5S 3H8, Canada}

\author[0000-0002-7351-6062]{Marc-Antoine Miville-Desch\^enes}
\affiliation{AIM, CEA, CNRS, Universit\'e Paris-Saclay, Universit\'e Paris Diderot, Sorbonne Paris Cit\'e, F-91191 Gif-sur-Yvette, France}

\author[0000-0003-2730-957X]{Naomi M. McClure-Griffiths}
\affiliation{Research School of Astronomy \& Astrophysics, Australian National University, Canberra ACT 2610 Australia}

\author[0000-0001-6846-5347]{Callum Lynn}
\affiliation{Research School of Astronomy \& Astrophysics, Australian National University, Canberra ACT 2610 Australia}

\author[0000-0003-0932-3140]{Andrea Bracco}
\affiliation{Laboratoire de Physique de l'Ecole Normale Sup\'erieure, ENS, Universit\'e PSL, CNRS, Sorbonne Universit\'e, Universit\'e de Paris, F-75005 Paris, France}

\author[0000-0001-7697-8361]{Luka Vujeva}
\affiliation{Niels Bohr International Academy, Niels Bohr Institute, Blegdamsvej 17, DK-2100 Copenhagen, Denmark}

\begin{abstract} 
% context heading (optional)
% aims heading (mandatory)
We develop a new method for spatially mapping a lower limit on the mass fraction of the cold neutral medium by analyzing the amplitude structure of $\hat T_b(k_v)$, the Fourier transform of $T_b(v)$, the spectrum of the brightness temperature of \HI\ 21\,cm line emission with respect to the radial velocity $v$.
This advances a broader effort exploiting 21\,cm emission line data alone (without absorption line data, $\tau$) to extract integrated properties of the multiphase structure of the \HI\ gas and to map each phase separately.
% methods heading (mandatory)
Using toy models, we illustrate the origin of interference patterns seen in $\hat T_b(k_v)$. Building on this, a lower limit on the cold gas mass fraction is obtained from the amplitude of $\hat T_b$ at high $k_v$.
% results heading (mandatory)
Tested on a numerical simulation of thermally bi-stable turbulence, the lower limit from this method has a strong linear correlation with the ``true" cold gas mass fraction from the simulation for relatively low cold gas mass fraction. At higher mass fraction, our lower limit is lower than the ``true" value, because of a combination of interference and opacity effects.
Comparison with absorption surveys shows a similar behavior, with a departure from linear correlation at $\NHI\gtrsim 3-5\times10^{20}$\,cm$^{-2}$.
Application to the DRAO Deep Field (DF) from DHIGLS reveals a complex network of cold filaments in the Spider, an important structural property of the thermal condensation of the \HI gas. 
Application to the HI4PI survey in the velocity range $-90 < v < 90$\,\kms\ produces a full sky map of a lower limit on the mass fraction of the cold neutral medium at 16\farcm2 resolution.
Our new method has the ability to extract a lower limit on the cold gas mass fraction for massive amounts of emission line data alone with low computing time and memory, pointing the way to new approaches suitable for the new generation of radio interferometers.
\end{abstract}

\keywords{ISM: structure – \,\,Methods: observational - data analysis}

\section{Introduction} \label{sec:intro}
%General statement
In the interstellar medium (ISM), thermal instability is thought to be the main process that leads to thermal condensation of the warm neutral phase, producing a thermally unstable lukewarm medium and a dense cold neutral medium \citep[CNM,][]{field_thermal_1965,field_cosmic-ray_1969,wolfire_1995,wolfire_2003,hennebelle_dynamical_1999,hennebelle_perrault_2000,audit_hennebelle_2005}. Understanding the properties of the multiphase neutral ISM is key to understanding this initial step leading to the atomic-to-molecular (HI-to-H$_2$) transition and the formation of molecular clouds \citep{lee_2012,lee_2015,stanimirovic_2014,burkhart_2015,pingel_2018,bialy_2019,wang_2020b,syed_2020,kalberla_2020,marchal_2021a,rybarczyk_2022,seifried_2022}.

%Surveys and challenge
%Over the past 
In recent decades, huge efforts have been made to map the 21\,cm emission of Galactic \HI \cite[e.g.,][]{taylor_2003,kalberla_2005,stil_2006,mcclure-griffiths_2009,martin_ghigls:_2015,hi4pi_2016,winkel_effelsberg-bonn_2016,beuther_2016,blagrave_dhigls:_2017,peek_2011,peek_2018,wang_2020a}, as well as \HI\ in nearby galaxies \citep[e.g.,][]{kim_1998,walter_2008,koch_2018,pingel_2022}.
%, and a large amount of data is now available. 
However, the spatial information about the multiphase and multi-scale nature of the neutral ISM contained in these large hyper-spectral position-position-velocity (PPV) cubes has remained elusive due to the difficulty in separating the emission from the different phases along each line of sight \citep[see][and references within]{marchal_2019}.

%Previous methods
%%Gaussian decomposition
In recent decades, methods aimed at phase separation of 21\,cm emission-only data 
have been developed to recover information about the physical properties of each phase. Among them, Gaussian decomposition algorithms have focused on modeling the contribution of each phase to \HI\ PPV cubes \citep{haud_2000,nidever_2008,lindner_2015,kalberla_2018,riener_2019,marchal_2019}. Through the use of spatial regularization of the Gaussian model, the code {\tt ROHSA} \citep{marchal_2019} was developed to model each phase with spatial coherence across velocities and allowed the thermal and turbulent properties of the WNM to be studied with emission line data only \citep{marchal_2021a}. However, these methods are often computationally expensive and difficult to apply to large data sets.

%%CNN
Furthering this ongoing effort,
\citet{murray_2020} have recently introduced a novel approach based on a 1D convolutional neural network (CNN) trained on a library of synthetic observations of emission line and absorption spectra of three-dimensional hydrodynamical simulations of the multiphase ISM \citep{kok_2013,kok_2014}. In contrast to Gaussian decomposition algorithms, \citet{murray_2020} focused on extracting the total cold gas mass fraction, and the correction to the optically thin estimate of \HI\ column density because of optical depth along each spectrum, losing information on the velocity field.
In this paper, although focusing on a similar goal (i.e., extracting the total cold gas mass fraction), 
we develop a new method by analyzing the amplitude structure of $\hat T(k_v)$, the Fourier transform (hereafter FT) of $T_b(v)$, the spectrum of the brightness temperature of \HI\ 21\,cm line emission with respect to the radial velocity $v$.

%Goal of this paper 
Previous studies performing spectral modeling in the $k_v$ domain, such as the Velocity Coordinate Spectrum (VCS) \citep[VCS,][]{lazarian_pogosyan_2000,lazarian_pogosyan_2006,lazarian_pogosyan_2008,lazarian_2009}, have shown that the power spectrum of spatially averaged 21\,cm data cubes contains information about the multiphase structure of the gas \citep{chepurnov_velocity_2006,chepurnov_velocity_2010,chepurnov_turbulence_2015}, but the VCS loses spatial information on the plane of the sky and precludes mapping out integrated properties of each phase.
Our new method using Fourier transformed spectra exploits valuable information on the amplitude in the 
$k_v$ domain along individual lines of sight and we obtain a spatial map of a lower limit on the contribution of the cold phase to the total line emission (i.e., the gas cold mass fraction).\footnote{A notebook that illustrates the method is available at \href{https://github.com/antoinemarchal/FFT-21cm}{https://github.com/antoinemarchal/FFT-21cm}. \label{foot:notebook}}

%Organization paper
The paper is organized as follows. 
In Section~\ref{sec:fourier}, we present the interference patterns seen in the FT of emission data from the 21 cm Spectral Line Observations of Neutral Gas with the VLA (21-SPONGE) survey \citep{murray_2015,murray_2018} and use toy models to understand their origins. Building on this, we develop a methodology to evaluate a lower limit on the mass fraction of gas below an arbitrary kinetic temperature.
Evaluation using a realistic numerical simulation of the neutral ISM is presented in Section~\ref{sec:simulation}.
In Section~\ref{sec:21-SPONGE-full}, we apply our new method to emission data alone from 21-SPONGE and compare our lower limits to the cold gas mass fraction obtained by \citet{murray_2018} by joint analysis of emission and absorption data.
Application to emission-line data for the DRAO Deep Field (hereafter DF) from the DHIGLS survey and the HI4PI survey in Sections~\ref{sec:DHIGLS} and \ref{sec:hi4pi}, respectively, shows that the information contained at high $k_v$ in PPK cubes\footnote{We will refer to the hyper-spectral data cube of 
Fourier transformed emission-line spectra in the $k_v$ domain as a position-position-$k_v$ (PPK) cube, akin to a PPV cube that describes the data in the velocity domain.} of Fourier transformed emission-line spectra is intimately related to the emission in small spatial scale features often seen in individual channel maps of PPV cubes.
A discussion and a summary are provided in Sections~\ref{sec:discussion} and \ref{sec:summary}, respectively.

\section{Fourier transform of 21\,cm data}
\label{sec:fourier}
\subsection{Fourier transform and $k_v = 0$ normalization}
The discrete FT of brightness temperature data $T_b(v)$ as a function of radial velocity $v$ is 
\begin{equation}
    \hat T_b(k_v) = \sum_v T_b(v) \, \exp \left( -2\,\pi\,j\,v\,k_v\right) \, dv \, ,
\end{equation}
where $j$ is the imaginary unit.
At $k_v=0$, $\hat T_b(0) = \sum_v T_b(v) \, dv$, which is the estimator of the total \HI\ column density in the optically thin limit, $N^*_{\rm HI}$, divided by $C=1.82243 \times 10^{18}$\,cm$^{-2}$\,(K km s$^{-1}$)$^{-1}$.
We use the information contained in the first mode to define the $k_v=0$ normalization of the FT data
\begin{align}
    \hat T'_b(k) = \hat T_b(k) / \hat T_b(0) \, ,
\end{align}
whose amplitude has values in the range [0, 1].

\subsection{A case from the 21-SPONGE survey}
\label{subsec:21-sponge}

\begin{figure}
  \centering
  \includegraphics[width=0.97\linewidth]{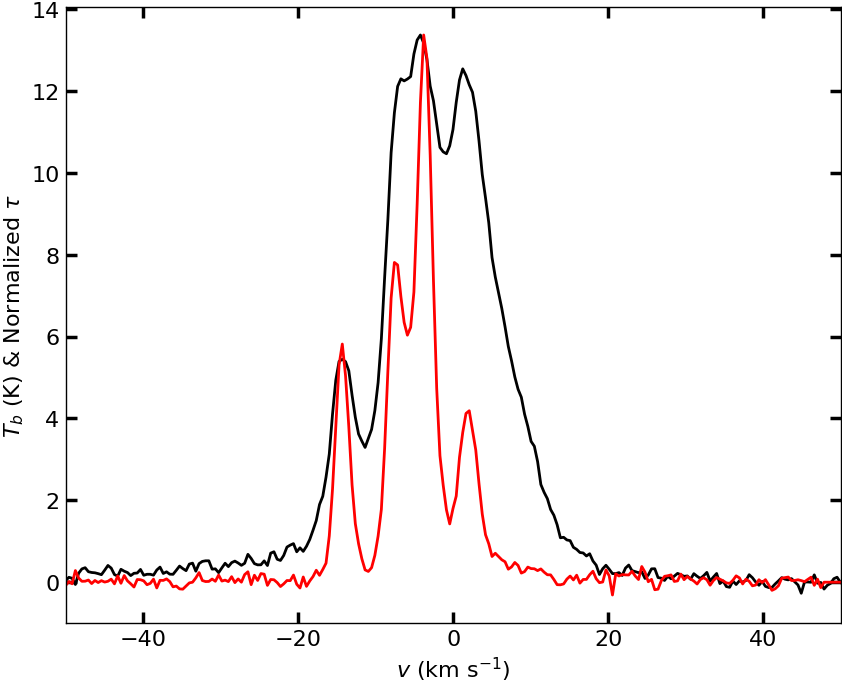}
  \caption{Brightness temperature $T_b$ of the absorption source J2232 from the 21-SPONGE survey (black). Its corresponding optical depth $\tau$, normalized at the peak of $T_b$, is shown in red.}
  \label{fig:diffraction_21Sponge_Tb}
\end{figure}

\begin{figure}
  \centering
  \includegraphics[width=\linewidth]{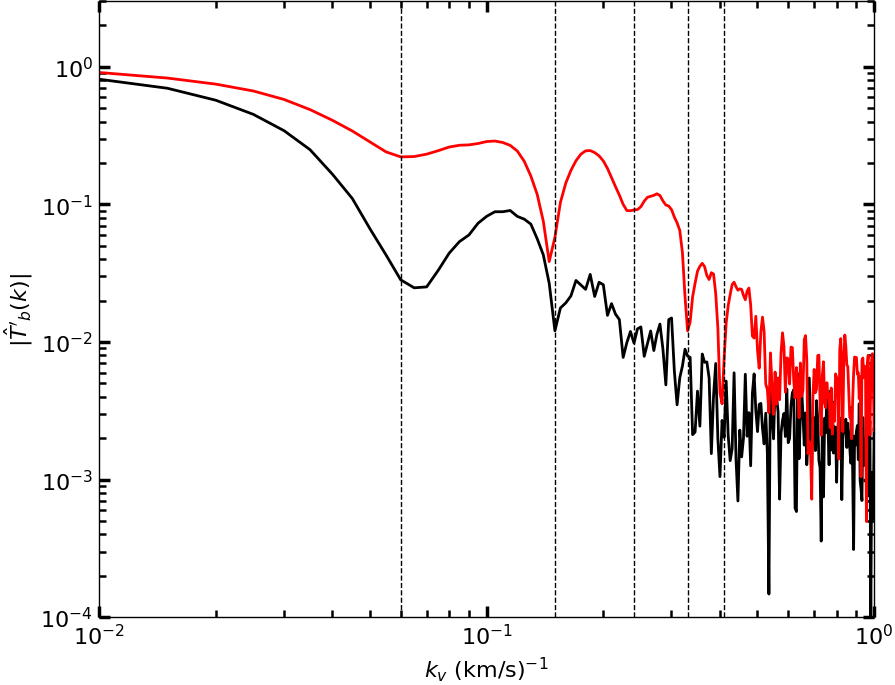}
  \caption{
  Amplitude of the normalized FT of $T_b$ (black) and $\tau$ (red) of the absorption source J2232 from the 21-SPONGE survey.
  The vertical dashed lines shows the position of dips observed in both amplitude spectra at similar $k_v$.}
  \label{fig:diffraction_21Sponge}
\end{figure}

%Example spectrum from 21-SPONGE
To illustrate the transformation of the data from the velocity domain to the $k_v$ domain, we used observations from 21-SPONGE.\footnote{Publicly available on the 21-SPONGE \href{https://dataverse.harvard.edu/dataset.xhtml?persistentId=doi:10.7910/DVN/BWFKL6}{Dataverse}} 
Specifically, we used observations of the 21-cm absorption-line optical depth $\tau(v)$ against the radio continuum source J2232 located at $(\alpha,\, \delta) = (22^{{\mathrm h}}$32$^{{\mathrm m}}36\fs4,\, 11\degree 43' 50\farcs9)$ or $(l,\, b) = (77\fdg 438,\, -38\fdg 582)$ obtained with the Karl G. Jansky Very Large Array (VLA).
The matching emission-line spectrum $T_b(v)$ was interpolated from off-source observations with the Arecibo telescope with angular resolution of about $3\farcm5$ and velocity resolution of 0.16 \kms (see details about the matching procedure in section 2.4 of \citealt{murray_2015}).
These complementary spectra are shown in Figure~\ref{fig:diffraction_21Sponge_Tb}, with the $\tau(v)$ spectrum scaled to the same maximum as $T_b(v)$ to highlight the relationship between the two.

Multiple components of cold gas were identified by \citet{murray_2018} using a joint multi-Gaussian decomposition of $T_b$ and $\tau$. Inferred optical depth, velocity, velocity dispersion, spin temperature, and column density of each component can be found in their table~5. The total mass fraction of cold gas that they detect in absorption along the line of sight against J2232 is 0.30$\pm$0.10 (see their table~2).

Figure~\ref{fig:diffraction_21Sponge} shows the amplitude of 
$k_0$ normalized Fourier transformed spectra of $T_b$ and $\tau$ on a logarithmic scale,
using the same color-code as in Figure~\ref{fig:diffraction_21Sponge_Tb}. 
Several ``dips" at similar $k_v$ are observed in the two amplitude spectra, marked by the vertical dashed lines.
Because the absorption spectrum ($\tau$) is revealing only cold gas along the line-of-sight, these dips observed in both the FT of $\tau$ and $T_b$ suggests that CNM gas plays an important role in shaping the FT of the emission spectrum as well, even though emission includes  contributions from \HI\ at all temperatures.
Note also that at high $k_v$, noise dominates the signal in both spectra. However, in the FT of $\tau$ the position of the dips can be traced unambiguously up to relatively high $k_v$ (low velocity dispersion), about $k_v\gtrsim0.4$\,(km/s)$^{-1}$.

\subsection{Understanding interference patterns in the amplitude of the FT spectrum}
\label{subsec:diffraction}

To understand the origin of the observed features in the Fourier transformed emission-line spectra, we used a toy model based on the mixing of multiple Gaussians with various amplitudes, velocities, and velocity dispersions (an arbitrary mixture of kinetic and turbulent broadening) to generate simplified mock observations of 21\,cm emission data.

\subsubsection{Toy model}
\label{subsubsec:toy-model}
The generative model is
\begin{equation}
  T_b\left(v, \theta\right) = \sum_{n=1}^{N} a_n \exp
  \left( - \frac{\left(v - \mu_n\right)^2}{2 \sigma_n^2} \right) \, ,
  \label{eq:model_gauss}
\end{equation}
where each of the $N$ Gaussians is parameterized by $\theta_n = \left(a_n , \, \mu_n , \, \sigma_n \right)$, with amplitude $a_{n} \geq 0$, mean velocity $\mu_n$, and velocity dispersion $\sigma_n$. 
Taking the Fourier transform and using the shift theorem, the normalized FT of the model is 
\begin{align}
    \hat T_b'(k_v,\theta'_n) &= \sum_{n=1}^{N} f_n \exp
  \left( - \frac{k_v^2}{2 \sigma_{k_v,n}^2} \right) \, e^{-2\,\pi\,j\,\mu_n\,k_v} \, ,
  \label{eq:fft-general}
\end{align}
where $\theta'_n = \left(f_n , \, \mu_n, \,  \sigma_{k_v,n} \right)$, $f_n=\sqrt{2\pi} a_n \sigma_n (C/ N^*_{\rm HI})$ is the mass fraction of each Gaussian $n$ (bounded between zero and unity) relative to the optically thin estimator of the total column density, and $\sigma_{k_v,n}=1/(2\pi \sigma_n)$.

\subsubsection{Single-Gaussian case}
\label{subsubsec:one-gaussian}

\begin{figure}
  \centering
\includegraphics[width=\linewidth]{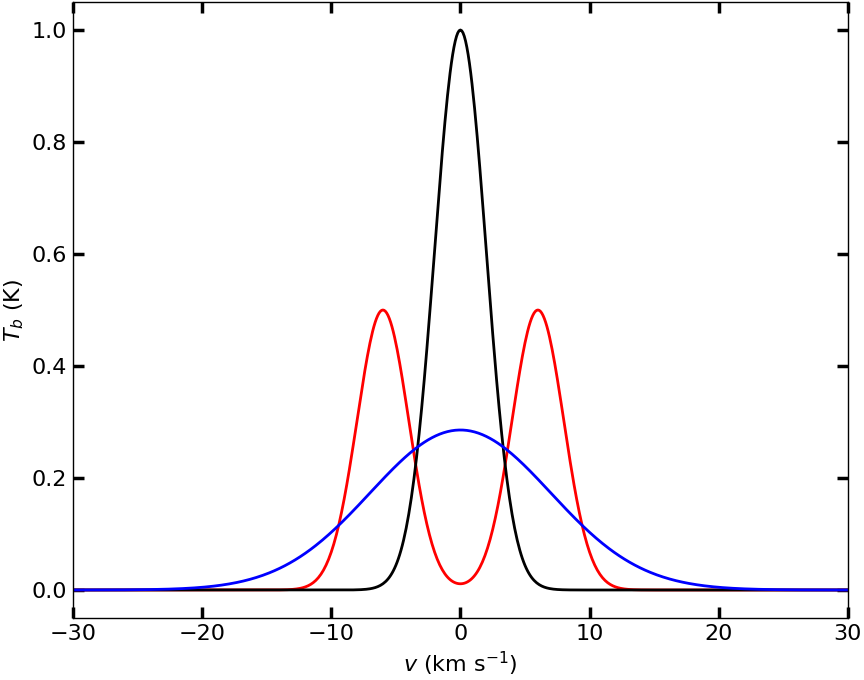}
  \caption{Brightness temperature spectrum $T_b$ of mock models. The black and blue curves show single-Gaussian models with no radial velocity (i.e., centered at $v=0$\,\kms) and $\sigma=2$\,\kms, and $\sigma=7$\,\kms, respectively. The red curve shows a double-Gaussian model with velocity separation $\Delta_v=12$\,\kms.}
  \label{fig:double_slit_Tb}
\end{figure}

\begin{figure}
  \centering
\includegraphics[width=\linewidth]{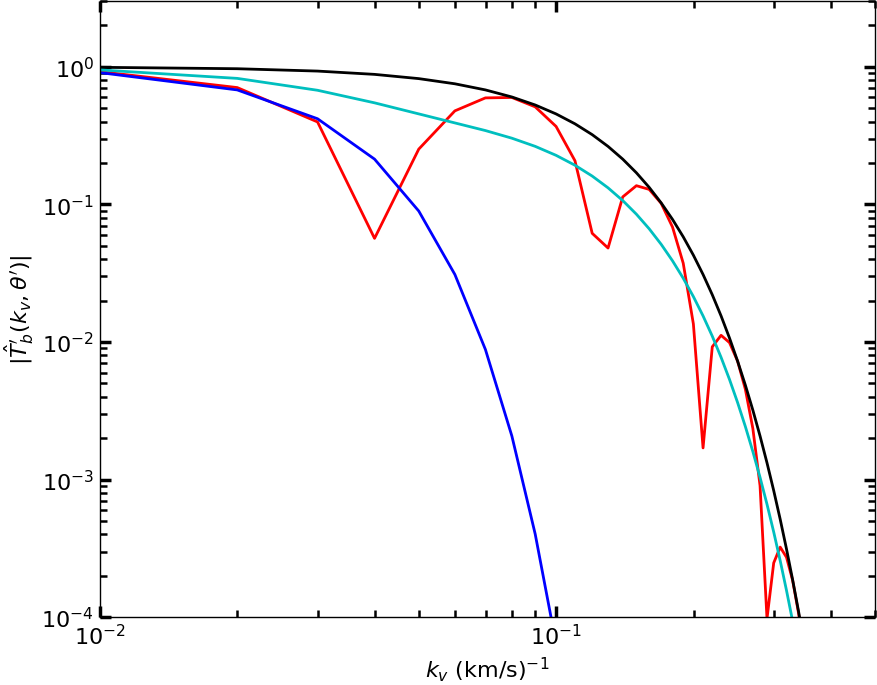}
  \caption{Normalized amplitude of the FT of the brightness temperature spectrum $T_b$ of toy models. Colors are the same as in Figure~\ref{fig:double_slit_Tb}, with the addition of the cyan curve for the model with two centered Gaussians in Section \ref{subsubsec:two-gaussians}.
  }
  \label{fig:double_slit}
\end{figure}
For a simplified model with $N=1$ the complex exponential that defines the phase of the signal, parametrized by $\mu$, simplifies when one considers the normalized amplitude of the FT,
\begin{align}
    |\hat T_b'(k_v,\theta'| &= f \exp \left( - \frac{k_v^2}{2 \sigma_{k_v}^2} \right) \, ,
    \label{eq:fft-single}
\end{align}
and can be fully parametrized by a single $f = 1$ and the corresponding $\sigma_{k_v}$, dropping the subscript $n = 1$ in Equation \ref{eq:fft-general}. The amplitude of the FT of a single Gaussian is also a Gaussian, centered on $k_v=0$ with a width inversely proportional to the velocity dispersion of the original Gaussian. 

%discussion
Therefore, a narrow (broad) Gaussian describing cold (warm) gas  $T_b\left(v, \theta\right)$ spectrum will appear broad (narrow) in its FT amplitude spectrum. 
This is illustrated in Figure~\ref{fig:double_slit_Tb} by the narrow ($\sigma=2$\,\kms) and broad ($\sigma=7$\,\kms) Gaussians with $\mu = 0$\,\kms\ (i.e., centered at $v=0$\,\kms) (black and blue lines, respectively), chosen to have the same area, representing the same column density of gas. Their corresponding normalized FT amplitude spectra are shown in Figure~\ref{fig:double_slit}. 

\subsubsection{Two centered Gaussians}
\label{subsubsec:two-gaussians}

Here the spectral model is the sum of the narrow (black) and broad (blue) Gaussians in Figure \ref{fig:double_slit_Tb}, though not shown there.  Thus, there is twice as much gas, half of which could be classified as cold; i.e., the cold gas mass fraction is 0.5.  The normalized amplitude of the FT is given by
\begin{align}
    |\hat T_b'(k_v,\theta')| &= f \sum_{n=1}^{N=2} \exp
  \left( - \frac{k_v^2}{2 \sigma_{k_v,n}^2} \right) \, ,
  \label{eq:fft-general-double}
\end{align}
where $f = f_1 = f_2 = 0.5$. 
%$\sigma_{k_v,1}=2$\,\kms, $\sigma_{k_v,2}=7$\,\kms, 
%
Because $\mu_1 = \mu_2$ there is no information from the phase of the FT (cf. Equation \ref{eq:double-slit}), and Equation~\ref{eq:fft-general-double} has the same functional form as the spectral model (i.e., a sum of two Gaussians). But now the broad component comes from the cold gas (black, $\sigma_{k_v,1}=0.079577$\,(\kms)$^{-1}$), while the narrow component originates from the warm gas (blue, $\sigma_{k_v,2}=0.022736$\,(\kms)$^{-1}$).
The normalized amplitude is plotted as the cyan curve in Figure \ref{fig:double_slit}.

%
%There is no information in the phase of the FT, hence the similarity.

\subsubsection{Analogy to Fraunhofer diffraction}
\label{subsubsec:fraunhofer}
It is useful to make an analogy to Fraunhofer diffraction.  When a plane wave passes through an aperture, the patterns on a distant screen can be calculated by the interference of secondary wavelets according to the Huygens-Fresnel principle.  The diffraction pattern calculated for a sharply bounded transparent aperture has characteristic oscillations, but the diffraction pattern calculated for an aperture with a Gaussian transmission profile is a smooth Gaussian.
This can be generalized to the case of multiple combined Gaussian transmission profiles of different widths but all centered at the same position.

Alternatively, in Fourier optics the Fraunhofer diffraction can be written as the FT of the aperture function, whose functional form again determines the shape of the diffraction pattern observed on the distant screen. Here, the brightness temperature $T_b(v)$ can be seen as analogous to a 1D aperture function and so the amplitude of its FT is analogous to the amplitude of the resulting 1D Fraunhofer diffraction pattern.

\subsubsection{A ``double-slit" analogy}
\label{subsubsec:double-slit}

When a plane wave passes through two slits, there is further interference of the secondary wavelets to be accounted for. The resulting diffraction can again be calculated with the Fourier transform.
This can be appreciated with a simple double-Gaussian model with $a_1 = a_2 = a = 0.5$, and $\sigma_{1} = \sigma_{2} = 2$\, \kms, and $\mu_1=-\mu_2=6$\,\kms. Note that this model, red in Figure \ref{fig:double_slit_Tb}, has the same column density of gas as the single Gaussian model (black) with the same $\sigma$.

Equation~\ref{eq:fft-general} simplifies to
\begin{align}
    |\hat T_b'(k_v,\theta')| &= f \, \exp
  \left( - \frac{k_v^2}{2 \sigma_{k_v}^2} \right) \, \left|\cos(\pi\,\Delta_v\,k_v)\right| \, ,
  \label{eq:double-slit}
\end{align}
where $f = 1$ and $\Delta_v=\mu_1-\mu_2=12$\,\kms\ is the velocity separation between the two Gaussians. 
This is the same as Equation \ref{eq:fft-single} for the single Gaussian, except now modulated by $|\cos(\pi\,\Delta_v\,k)|$.
The normalized amplitude of its FT is shown by the red curve in Figure \ref{fig:double_slit}. The spectral resolution was set to 0.1\,\kms.
This modulation affects the Gaussian function only destructively and so the single-Gaussian case (black) forms an envelope to the amplitude of the FT of the double-Gaussian model (red), as is familiar in the Fraunhofer diffraction pattern of any double slit experiment.

\subsubsection{Multiple Gaussians}
\label{subsubsec:multiple-gaussians}

\begin{figure}
  \centering
  \includegraphics[width=\linewidth]{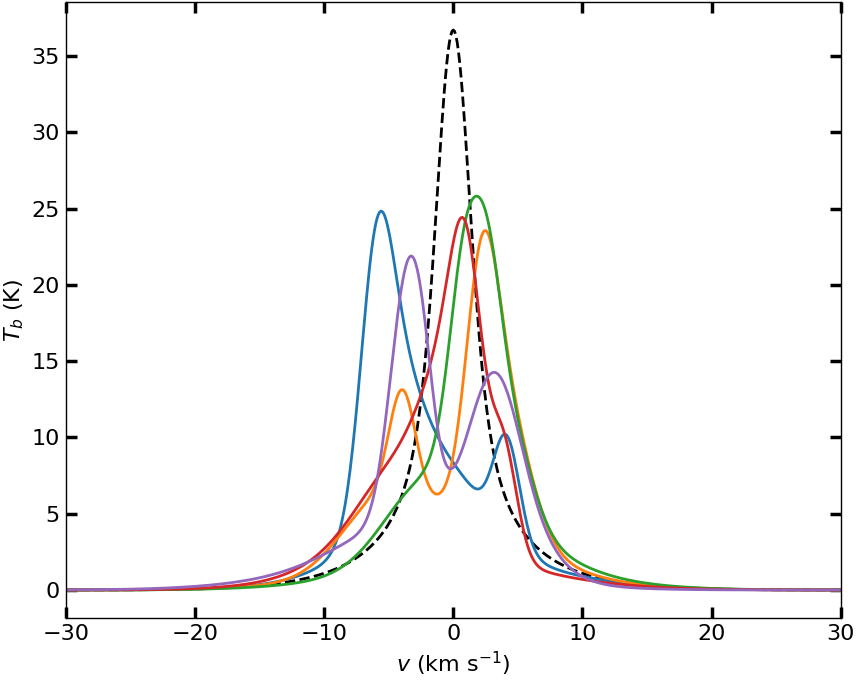}
  \caption{Brightness temperature spectrum $T_b$ of six-Gaussian toy models of a multi-Gaussian component multiphase gas. The black dashed curve shows $T_b$ of the baseline model with $\mub_n=\bm{0}$, and randomly selected $\ab_n$ and $\sigmab_n$ (see text). Colored lines show $T_b$ of five more models with the same $\ab_n$ and $\sigmab_n$ but with $\mub_n$ chosen randomly in the range $-6$\,\kms\ $< \mu < 6$\,\kms.}
  \label{fig:diffraction_toy_model_Tb}
\end{figure}

\begin{figure}
  \centering
\includegraphics[width=\linewidth]{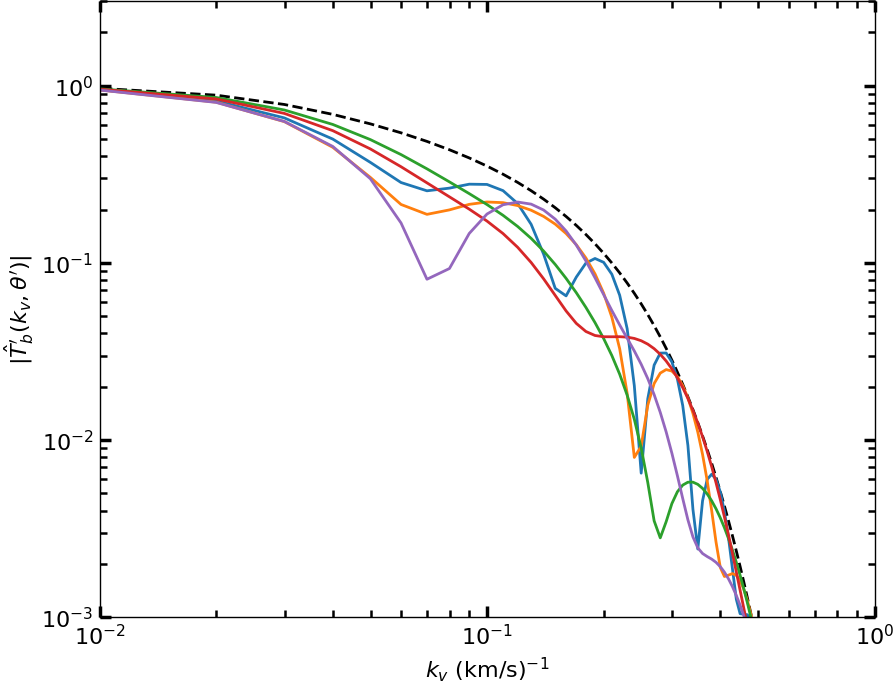}
  \caption{Normalized amplitude of the FT of the brightness temperature spectrum $T_b$ of toy models of a multi-Gaussian component multiphase gas. Annotations are as in Figure~\ref{fig:diffraction_toy_model_Tb}.
   }
  \label{fig:diffraction_toy_model}
\end{figure}

Using Equation~\ref{eq:model_gauss}, we can generate more realistic synthetic spectra of a turbulent fluid with multiple components of various widths representing multiple thermal phases. Specifically, we built six-Gaussian models with one Gaussian assigned to the velocity dispersion range 6\,\kms\ $< \sigma < 8$\,\kms\ (about 4400\,K $< T_{k,\rm max} < 7800$\,K), two in the range 3 to 6\,\kms\ (about 1100\,K $< T_{k,\rm max} < 4400$\,K), and three in the range 0.7 to 2\,\kms\ (about 70\,K $< T_{k,\rm max} < 500$\,K). 
The six amplitudes $\ab_n$ were chosen randomly, in the range 0 to 16\,K.

A baseline model was made with all components centered at $\mu_n=0$ to produce a non-destructive reference.
Next, five more models were built with six velocities $\mub_n$ randomly chosen in the velocity range $-6$\,\kms\ $< \mu < 6$\,\kms, to be different for each of these five models.
However, the six models share the same amplitudes $\ab_n$ and velocity dispersions $\sigmab_n$ and hence have the same mass fractions associated with each of the six individual Gaussians. 

Figure~\ref{fig:diffraction_toy_model_Tb} shows the brightness temperature of all six models for a case with cold gas mass fraction (the three Gaussians with $\sigma < 3$\,\kms) about 0.5.
The spectral resolution was again set to 0.1\,\kms. The black dashed curve shows the baseline model with $\mub_n=0$. The five colored curves show models with varying $\mub_n$.
Figure~\ref{fig:diffraction_toy_model} shows their FT amplitude spectra, using the same color code. As expected from the single and double-Gaussian models, interference patterns are seen in the five models with $\mub_n \neq 0$, resulting from the multiple peaks in their $T_b$ spectra. 
As in the double-Gaussian model, interference appears only destructively and the $\mub_n=0$ baseline model provides an envelope for the other five models, regardless of the specific patterns produced by the random positions of each Gaussian.

%\subsection{A lower limit on the mass fraction of gas with selected maximum velocity dispersion (maximum kinetic temperature)}
\subsection{A lower limit on the cold gas mass fraction}
\label{subsec:lower-limit}

\begin{figure}
  \centering
  \includegraphics[width=\linewidth]{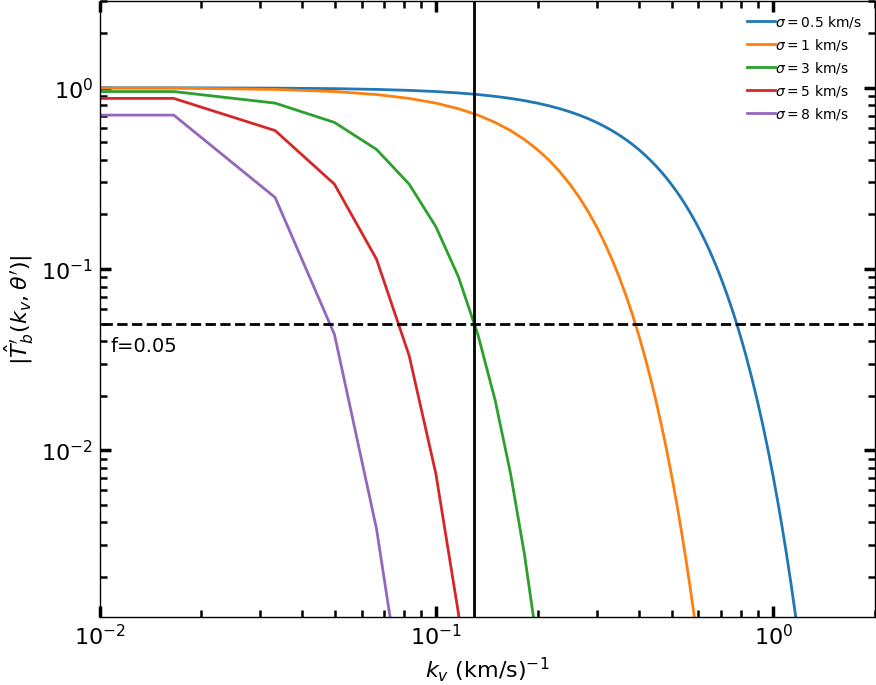}
  \caption{Normalized amplitude of the FT spectrum of single Gaussians with various width $\sigma$. The horizontal dashed line denotes the mass fraction $f=0.05$ and the vertical line its corresponding $k_v$ for the Gaussian with $\sigma =3$\kms.
  }
  \label{fig:single_gaussian_fft}
\end{figure}

Building on our understanding of patterns produced in the FT amplitude spectrum by the multi-component nature (phase and clouds/turbulent structures) of 21\,cm data, here we develop a new method to set a lower limit on the mass fraction of gas with an arbitrary maximum velocity dispersion (maximum kinetic temperature) in PPV cubes of \HI\ data. Specifically, we focused our analysis on the cold gas mass fraction.

Figure~\ref{fig:single_gaussian_fft} shows the normalized amplitude of the FT of single-Gaussian models with $0.5 < \sigma < 8$\,\kms. 
Gaussians with larger $\sigma$ appear narrower in the $k_v$-domain. 
In general, the presence of cold gas (narrow spectral lines) is detectable by the fact that the amplitude of the FT remains high at relatively large $k_v$.  Furthermore, if there is warm gas contributing to the spectrum, it will not be detectable in the amplitude of the FT at high $k_v$, but it will affect the normalization.  Therefore, what we can estimate at high $k_v$ is a lower limit to the cold gas mass fraction of the emitting gas.
%
%The y axis can be seen as the contribution to the gas mass fraction as a function of $k_v$. 

%For example, 
Focusing on colder gas, at the vertical black line $k_v=k_{v,\rm lim}=0.12$\,(\kms)$^{-1}$ gas characterized by a Gaussian with $\sigma = \sigma_{\rm lim}=3$\kms\ can be detected in the FT, though only at a threshold fractional level $t = t_{\rm lim} = 0.05$ of its true total mass. The value of the normalized amplitude is thus an underestimate of the total gas present in an amount dependent on $k_v$ and $\sigma$.

\begin{figure}
  \centering
  \includegraphics[width=\linewidth]{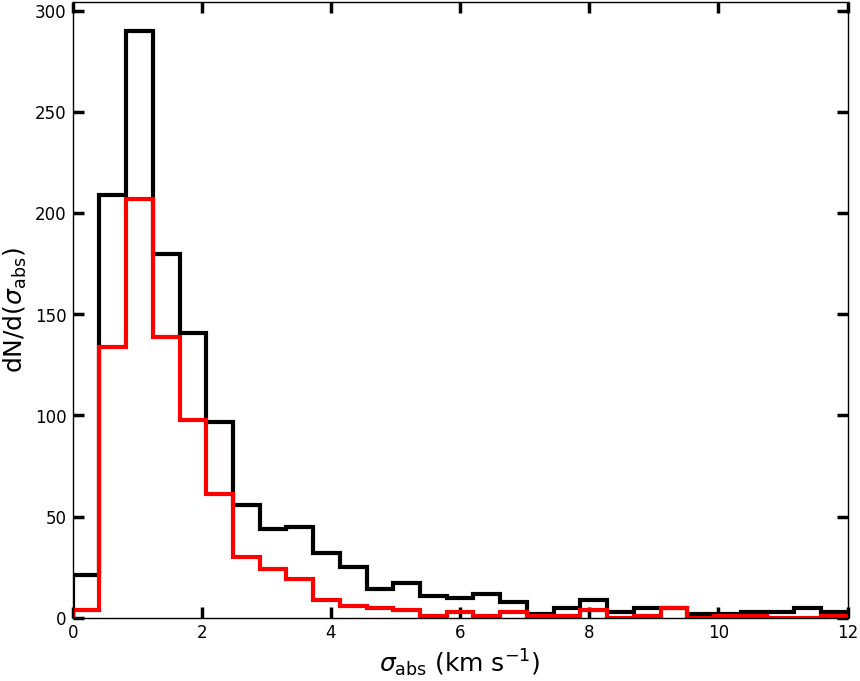}
  \caption{Histogram of $\sigma$ of absorption line components in the BIGHIHAT catalog (black).  The histogram for the CNM subset, defined as having $T_s < 250$ K, is also plotted (red).
  }
  \label{fig:sigmapdf}
\end{figure}

Without measurements of the spin temperature $T_s$, which is close to the kinetic temperature $T_k$ in cold gas, we need to consider line widths in the $T_b$ spectra. Using the approximation
\begin{align}
    T_{k, \rm max} \simeq 121 \, \sigma_{\rm lim}^2 \, ,
\end{align}
in units of K, $\sigma = 3$\kms\ corresponds to gas with a maximum kinetic temperature of about 1000 K.  
In reality, accounting for turbulent broadening, $T_k$ is probably much less.
%Warmer gas with broader $T_b$ spectra is more difficult to detect at a given $k_v$.
This can be assessed by looking at the data in the BIGHICAT meta-catalog compiled by \citet{mcclur23} (Section \ref{sec:validation-bighicat}). Figure \ref{fig:sigmapdf} shows a histogram of $\sigma$ for all Gaussian components identified in absorption with closely matching profiles in emission.  These analyses typically use $T_s < 250$ K to define CNM, not the line width (because of turbulent broadening). With that definition, we overplot the histogram for the subset that are CNM components. This shows that we want an estimator that detects gas with $\sigma$ up to a few \kms. Most of this gas will be CNM and any unstable LNM with be suppressed.

As first introduced in Section \ref{subsubsec:double-slit}, destructive interference patterns caused by multiple peaks in the $T_b$ spectra can reduce the amplitude of the FT at $k_v=k_{v,\rm lim}$, lowering detectability
of the contribution to the $T_b$ signal coming from Gaussians with $\sigma\leq \sigma_{\rm lim}$. Note that because of the modulation, higher values of the amplitude might occur at higher $k_v>k_{v,\rm lim}$, improving detectability, though still providing a lower limit on the gas mass fraction.

Following this discussion,
%reasoning, 
we define the estimator of the lower limit on the cold gas mass fraction to be
\begin{align}
   f^{k_{v,\rm lim}}_{\rm low} = \max \left[ \hat T'_b(k_v>k_{v,\rm lim}) \right] \, .    \label{eq:mass_fraction}
\end{align}
Except where specified in the appendices, in the rest of this paper we have evaluated this estimator of the cold gas mass fraction with $k_{v,\rm lim}=0.12$\,(\kms)$^{-1}$. 
%($\sigma_{\rm lim}=3$\kms, $t_{\rm lim} = 0.05$). 
For compactness, this is denoted
$f^{0.12}_{\rm low}$.
For other applications based on the maximum kinetic temperature one wishes to probe in the data (e.g., probing the structure of cold gas and unstable gas simultaneously), $k_{v,\rm lim}$ can be chosen judiciously following the methodology previously described pairing $\sigma_{\rm lim}$ and $t_{\rm lim}$. 

\subsubsection{Assessing \lowercase{$f^{0.12}_{\rm low}$} for toy models}
\label{subsubsec:limit-toymodels}

First, we evaluated $f^{0.12}_{\rm low}$
for the instructive cases in Figure \ref{fig:double_slit}.
In the case of a simple spectrum with a single peak, the amplitude of the FT (blue, black) gives a direct quantitative estimate. 
Our estimator evaluates as $1 \times 10^{-7}$ for the warm gas (blue); i.e., warm is not detectable. %
For the cold gas (black) the estimate is 0.26 for the selected $k_{v,\rm lim}$, thus detectable but lower than the true value of the cold gas mass fraction, which is 1.0. In this simple model, only for $\sigma$ even smaller than 2\,\kms\ does this estimator approach unity (Figure \ref{fig:single_gaussian_fft}).
For the mixture of warm and cold gas (cyan), the warm gas is not detectable directly but affects the normalization and we find 0.13 compared to the true value 0.5.
For the double Gaussian model (red), which has the same amount of cold gas as the black model, the amplitude of the FT is smaller than the black envelope because of the modulation, by a factor depending on the (range of) $k_v$ examined. Because of this modulation, the estimate of the cold gas mass fraction in more complex spectra can be even lower than the true value; for the red curve we find 0.14, compared to 0.26 for the black curve and the true 1.0.

\begin{figure}
  \centering
  \includegraphics[width=\linewidth]{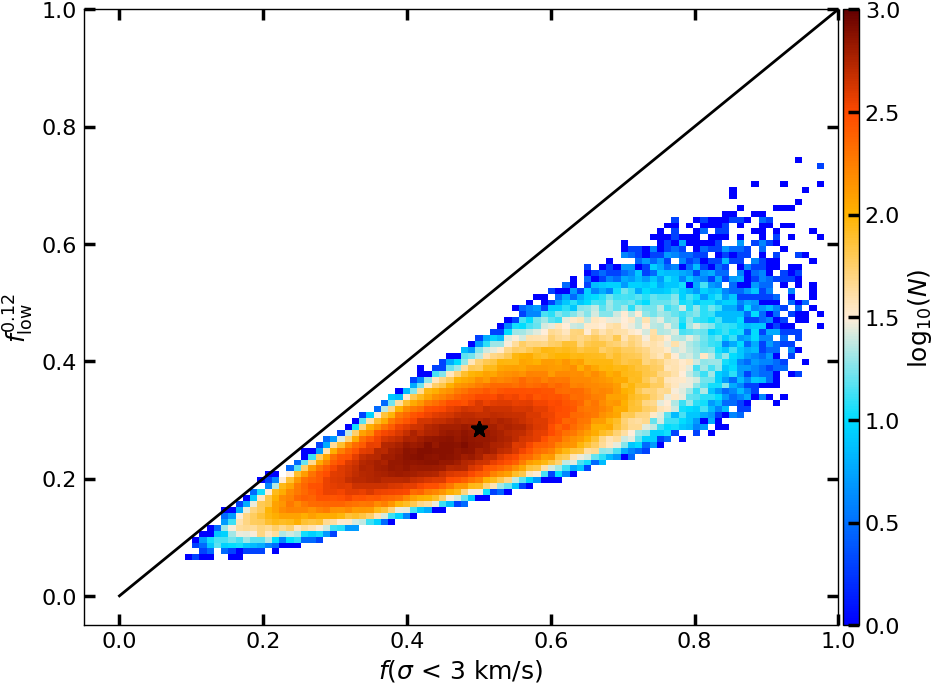}
  \includegraphics[width=\linewidth]{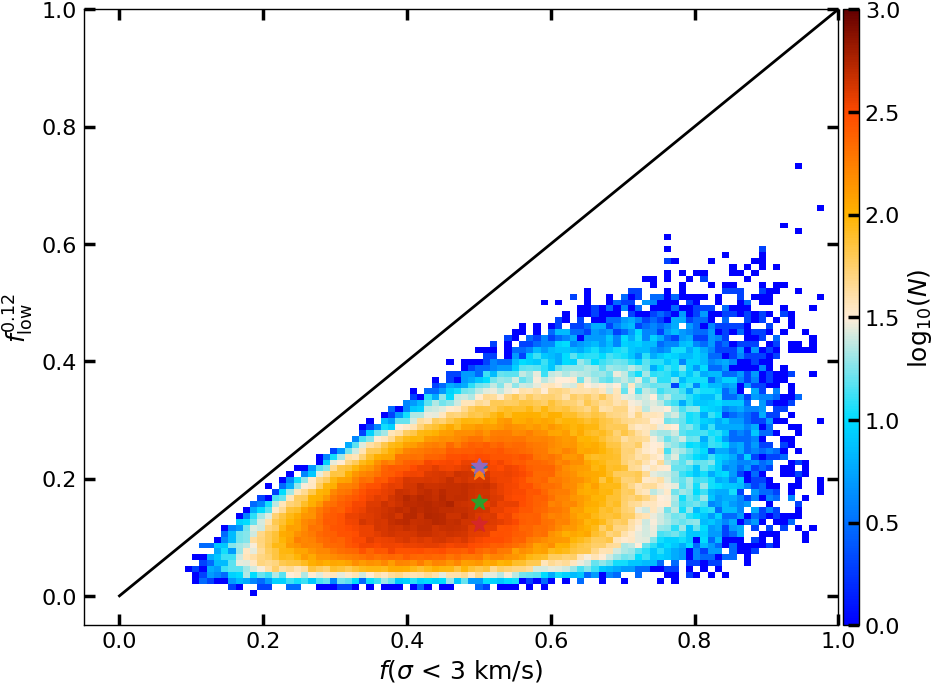}
  \caption{2D histograms of the estimator $f^{0.12}_{\rm low}$ and the actual $f(\sigma < 3 $\kms) for $2^{18}$ six-Gaussian toy models with random amplitudes and dispersions in the parameter ranges used in Figure~\ref{fig:diffraction_toy_model_Tb} (see text). 
  Top: spectra modified to have all $\mub_n = 0$. The black star is for the baseline model in Figure \ref{fig:diffraction_toy_model_Tb}.
  Bottom: spectra retaining random $\mub_n$ in the range $-6$\,\kms\ $< \mu < 6$\,\kms.   
  The colored stars are for the other five spectra that have interference in Figure~\ref{fig:diffraction_toy_model}, which lowers the estimator.}
  \label{fig:heatmap_toy_CNM_fftcut}
\end{figure}

Similarly, for the more realistic six-Gaussian models in Figure~\ref{fig:diffraction_toy_model}, we have evaluated $f^{0.12}_{\rm low}$.
Our estimator evaluates as 0.26 for the baseline model (black line in Figure~\ref{fig:diffraction_toy_model}), about half of the true input cold gas mass fraction, which is about 0.5.
For the related five models with random $\mub_n$ in the range $-6$\,\kms\ $< \mu < 6$\,\kms\ (colored lines), our estimator fluctuates due to interference but is lower than the value inferred from the baseline model.

We have expanded this six-Gaussian experiment to a sample of $2^{18}$ spectra with randomly selected amplitudes and dispersions (hence cold gas mass fractions) and velocities $\mub_n$ in the same parameter ranges described in connection with Figure~\ref{fig:diffraction_toy_model_Tb}.
The top panel of Figure~\ref{fig:heatmap_toy_CNM_fftcut} shows the 2D histogram of the estimator $f^{0.12}_{\rm low}$ vs.\ the actual $f(\sigma < 3 $\kms) for 
modified spectra resulting from setting all $\mub_n = 0$, like the baseline model in Figure \ref{fig:diffraction_toy_model_Tb}, which is shown here as the black star. 
For a constant $f(\sigma < 3 $\kms), there is a vertical spread of $f^{0.12}_{\rm low}$ though still consistent with being a lower limit.
This illustrates well that our estimator is sensitive to the specific distribution of the underlying amplitudes and dispersions for all phases.

The bottom panel is based on the original $2^{18}$ spectra including the non-zero $\mub_n$.
The colored stars are for the other five spectra in Figure~\ref{fig:diffraction_toy_model}.
This illustrates again that destructive interference lowers $f^{0.12}_{\rm low}$ but our estimator is still aptly called a lower limit.
We also note that even with interference, $f^{0.12}_{\rm low}$ retains a good correlation with $f(\sigma < 3 $\kms).

\section{Assessing \lowercase{$f^{0.12}_{\rm low}$} using a numerical simulation} \label{sec:simulation}

Because of the simplicity of the toy models, it is unclear if $f^{0.12}_{\rm low}$ also retains information about the spatial structure of cold gas in a realistic situation.  This is investigated here with the high-resolution hydrodynamical simulation of thermally bi-stable turbulence performed by \citet{saury_2014} using the three-dimensional radiation hydrodynamics code HERACLES \citep{gonzales_2007}.

\subsection{Numerical simulation}

%To test our novel FT-based estimate of the lower limit on the cold gas mass fraction in \HI\ emission lines, 
%we 
%
In HERACLES, heating and cooling processes were implemented based on the prescription of \citet{wolfire_1995,wolfire_2003}, including cooling by [CII], [OI], recombination onto positively charged interstellar grains, and Lyman $\alpha$, and heating dominated by the photo-electric effect on small dust grains and PAHs. In their simulation suite, \citet{saury_2014} considered a spatially uniform radiation field with the spectrum and intensity of the Habing field \citep[$G_0$/1.7, where $G_0$ is the Draine flux:][]{habing_1968,draine_1978}.

Specifically, we used part of their 1024N01 simulation ($1024^3$ pixels and a physical size of the box of 40\,pc) characterized by an initial volume density $n_0=1$\,cm$^{-3}$, a large-scale velocity $v_S=12.5$\,km\,s$^{-1}$ and a spectral weight $\zeta = 0.2$ that controls the modes of the turbulent mixing (here a majority of compressible modes). 
The initial density corresponds to the typical density of the WNM before thermal instability and condensation, and the large-scale velocity represents the amplitude given to the field that generates large-scale turbulent motions in the box.
A stationary state was reached after running the simulation for about 16\,Myr out of a total of 40\,Myr. The snapshot used here was taken after 16\,Myr \citep{saury_2014}.

\begin{figure}
  \centering
  \includegraphics[width=\linewidth]{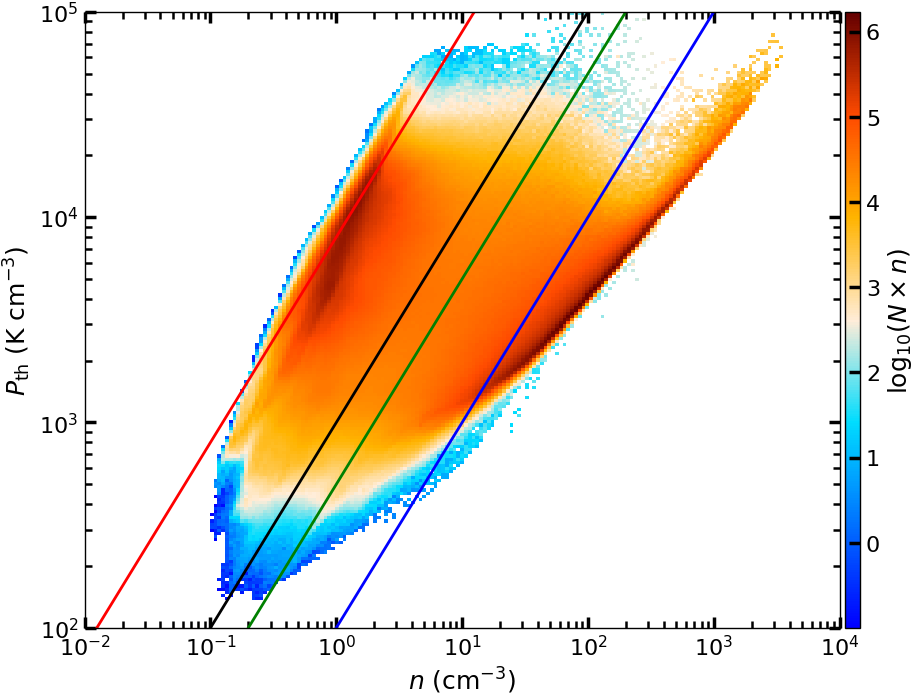}
  \caption{Pressure vs.\ volume density histogram weighted by volume density. The red, black, green, and blue lines indicate isothermal curves at 8000\,K, 1000\,K, 500\,K, and 100\,K, respectively.}
  \label{fig:P-n}
\end{figure}

Figure~\ref{fig:P-n} shows the 2D histogram of thermal pressure $P_{\rm th}$ and volume density $n$ for cells in the numerical simulation, weighted by $n$. The red, black, green, 
%magenta, 
and blue lines indicate isothermal curves at 8000\,K, 1000\,K, 500\,K, 
%240\,K, 
and 100\,K, respectively.

To explore the effect of opacity we concentrate our analysis on cells in a $512\times512\times 1024$\, pixel region with a large range of cold gas mass fraction. 

\subsection{21\,cm line synthetic observations}
\label{subsec::synthetic_obs}

Following \citet{marchal_2019}, we generated synthetic 21\,cm \HI\ observations both in the optically thin limit and considering the full radiative transfer, which is sensitive to the optical depth of the line. We used an effective velocity resolution of 0.8 km s$^{-1}$ in the velocity range $-40 < v < 40$\,km\,s$^{-1}$.

In a given cell at spatial coordinates $\rb$, the turbulent velocity field of the gas along the line of sight (taken to be the z axis) is $v_z(\rb,z)$.
The 1D distribution function of the z-component of the Maxwellian velocity
distribution in that cell is given by $\phi_{v_z}(\rb,z)$, a normal distribution (Gaussian) shifted by $v_z(\rb,z),$
\begin{align}
  \phi_{v_z}(\rb,z)dv' = \frac{1}{\sqrt{2\pi}\Delta(\rb,z)} \, \times \, \exp \left( - \, \frac{(v' \, - \,
    v_z(\rb,z))^2}{2\Delta^2(\rb,z)} \right)dv' \, ,
  \label{eq::Maxell_Boltzmann_law}
\end{align} \\
where $\Delta(\rb,z) = \sqrt{k_B T(\rb,z)/m_{\rm H}}$ is the thermal broadening of the 21\,cm line, $m_{\rm H}$ is the mass of the hydrogen atom, $k_B$ is the Boltzmann constant, and $T(\rb,z)$ is the gas temperature.

The general case for the computation of the 21\,cm  brightness temperature $T_b(v_z, \rb)$ is based on the following radiative transfer equation: 
\begin{align}
        T_b(v_z, \rb) = \sum_z T_s(\rb,z) \, \left[ 1 - e^{- \tau(v_z,\rb,z)} \right] \,
        e^{- \sum_{z'<z} \tau(v_z,\rb,z')}, 
        \label{eq:Tb_thick}
\end{align}
where $\tau(v_z,\rb,z)$ is the optical depth of the 21-cm line across the cell defined as
\begin{align}
        \tau(v_z,\rb,z) = \frac{1}{C} \, \frac{\rho(\rb,z) \, \phi_{v_z}(\rb,z)}
    {T_s(\rb,z)} \, dz \, ,
\end{align}
where $\rho(\rb,z)$ is the gas density. %
In this representation, emission by gas in the cell at $z$ is absorbed by gas in foreground cells at positions $z' < z$. 

A complete derivation of $T_s(\rb,z)$ requires determination of the level populations of the hyper-fine state of the 21\,cm line \citep[see e.g., section 2.3 in][]{kok_2014}. The three dominant mechanisms are: collisional transitions, direct radiative transitions by 21\,cm photons, and indirect radiative transitions (e.g., Ly$\alpha$ scattering). 
To simplify our calculation of the opacity, we assume here that $T_s(\rb,z)=T(\rb,z)$, motivated by the fact that collisions dominate the transition between levels in the cold dense phase. 
In the warmer phase $T_s$ might depart from $T$ if the rate of collisional transitions is insufficient to thermalize the gas, resulting in low values of $T_s$ in the range 1000--4000\,K \citep{liszt_2001}. However, the detection of a widespread warm component in 21-SPONGE studies with $\left<T_s\right>=7200^{+1800}_{-1200}$\,K \citep{murray_2014} or higher,\footnote{Following spectral line modeling, \citet{murray_2018} stacked
residual absorption in bins of residual emission and detected
a significant absorption feature with a harmonic mean
 $T_s$ about $10^4$\, K.} 
suggests that 
%Ly$\alpha$ scattering is more important than predicted by \citet{liszt_2001}, 
additional excitation mechanisms such as resonant Ly$\alpha$
scattering are present,
making $T_s$ close to the kinetic temperature of the gas.

In the optically thin limit where the opacity is negligible (i.e., $\tau(v_z,r,z)<<1$ everywhere), the 21\,cm brightness temperature is
\begin{align}
        T_b^*(v_z,\rb) \, dv' =
        \, \frac{1}{C} \int_0^L dz
    \, \rho(\rb,z) \, \phi_{v_z}(\rb,z) \, dv' \, ,
    \label{eq:optical_thin}
\end{align}
where $L$ is the total depth of atomic gas along the line of sight. Of course, $T_b^*(v_z,\rb)$ can be calculated even when the gas is not optically thin, and the difference of $T_b(v_z,\rb)$ with respect to this reveals the effect of opacity.

We generated two additional cubes, $T_b(v_z=0,\rb)$ and $T_b^*(v_z=0,\rb)$, both with the turbulent velocity field fixed at 0\kms\ for each cell in the simulation.
These four cubes were used to explore the independent effects of opacity and of interference patterns induced by the turbulent velocity field on the Fourier transformed 21\,cm emission line.
%Example spectrum for the 4 cases 
\begin{figure}
  \centering
  \includegraphics[width=\linewidth]{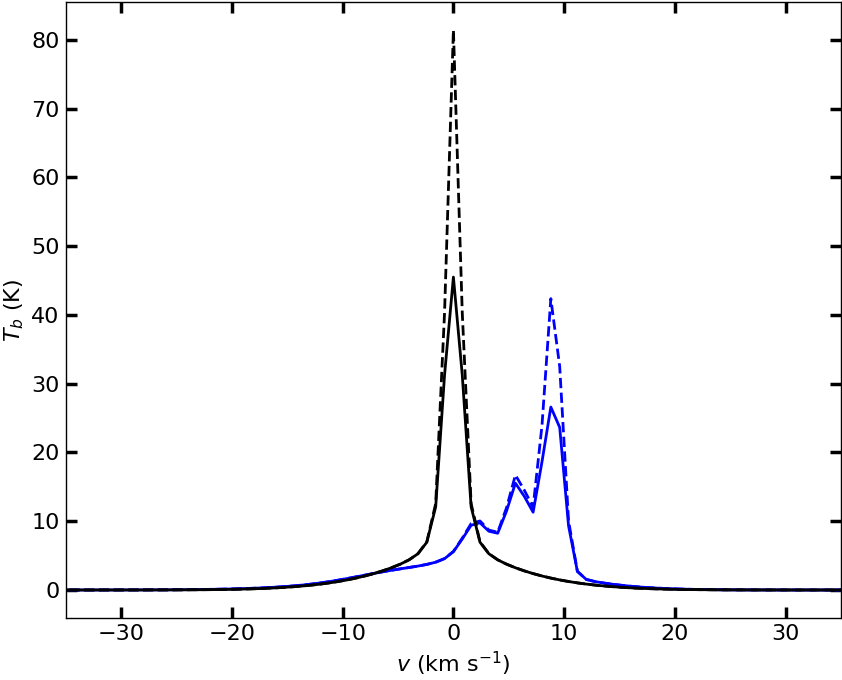}
  \caption{
  Brightness temperature spectra of a randomly selected line of sight within the four variants of the simulated PPV cube: $T_b^*(v_z=0,\rb)$ (dashed black), $T_b^*(v_z,\rb)$ (dashed blue), $T_b(v_z=0,\rb)$ (black), and $T_b(v_z,\rb)$ (blue).
  }
  \label{fig:Tb_random_4_cases_Saury}
\end{figure}
Figure~\ref{fig:Tb_random_4_cases_Saury} shows brightness temperature spectra of a randomly selected line of sight within these four variants of the simulated PPV cube to illustrate the shape of the simulated spectra and the effects of opacity and velocity field on the line.

%Fit and pearson correlation coefficient.
\begin{deluxetable}{lcccc}
\tablecaption{Summary statistics of the correlation between $f^{0.12}_{\rm low}$ and $f(T < T_{k,\rm max}$) for the four cases presented in Figures~\ref{fig:fcnm} and \ref{fig:heatmap_CNM_fftcut}}
\label{tab:statistic}
\tablewidth{0pt}
\tablehead{
\nocolhead{} & \colhead{$T_b^*(v_z=0,\rb)$} & \colhead{$T_b^*(v_z,\rb)$} & \colhead{$T_b(v_z=0,\rb)$} & \colhead{$T_b(v_z,\rb)$}
}
\startdata
{Pearson $r_p$} & 0.98 & 0.89 & 0.98 & 0.88 \\
{Slope} & 0.82 & 0.64 & 0.59 & 0.46 \\
{Intercept} & $-0.04$ & $-0.04$ & 0.01 & $-0.02$ \\
{$\left<cc(k)\right>$} & 0.88 & 0.75 & 0.83 & 0.71 \\
\enddata\end{deluxetable}

%%%%%%%%%%%%%%%%%%%%%%%%%%%%%%%%%%%%%%
% NOTE TO typesetter: ideally this figure and the following are on the same page
%%%%%%%%%%%%%%%%%%%%%%%%%%%%%%%%%%%%%%
\begin{figure}
  \centering
  \includegraphics[width=0.88\linewidth]{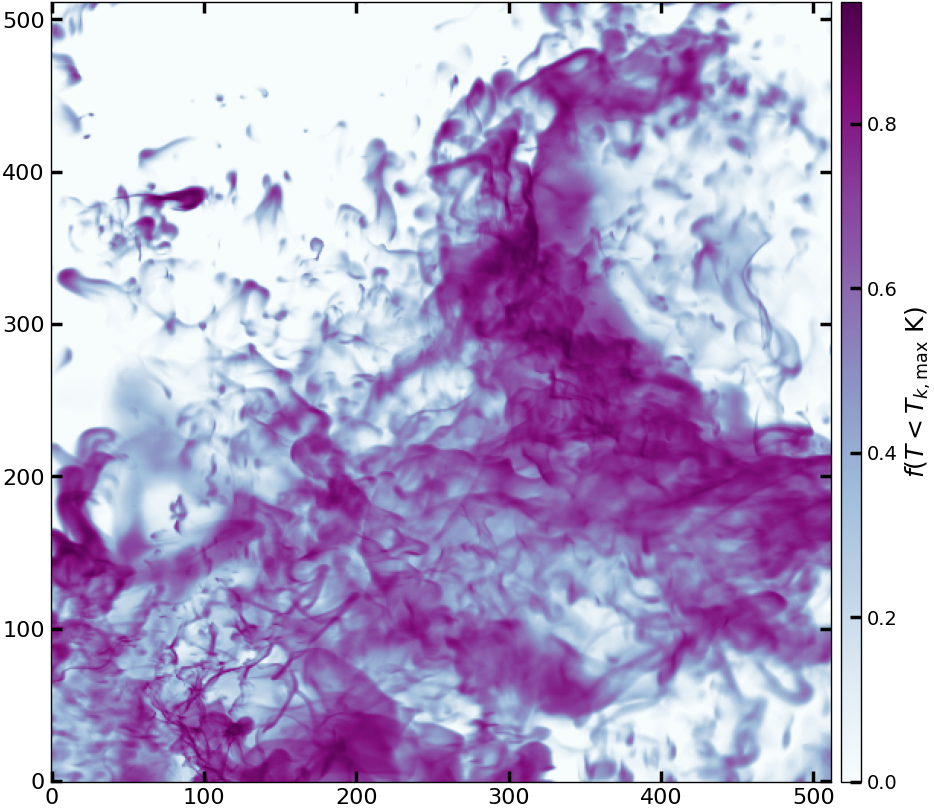}
  \caption{Cold gas mass fraction with $T < T_{k,\rm max}$ obtained directly from the simulated density field.}
  \label{fig:fcnm_True_saury}
\end{figure}

\begin{figure*}
  \centering 
  \includegraphics[width=0.42\linewidth]{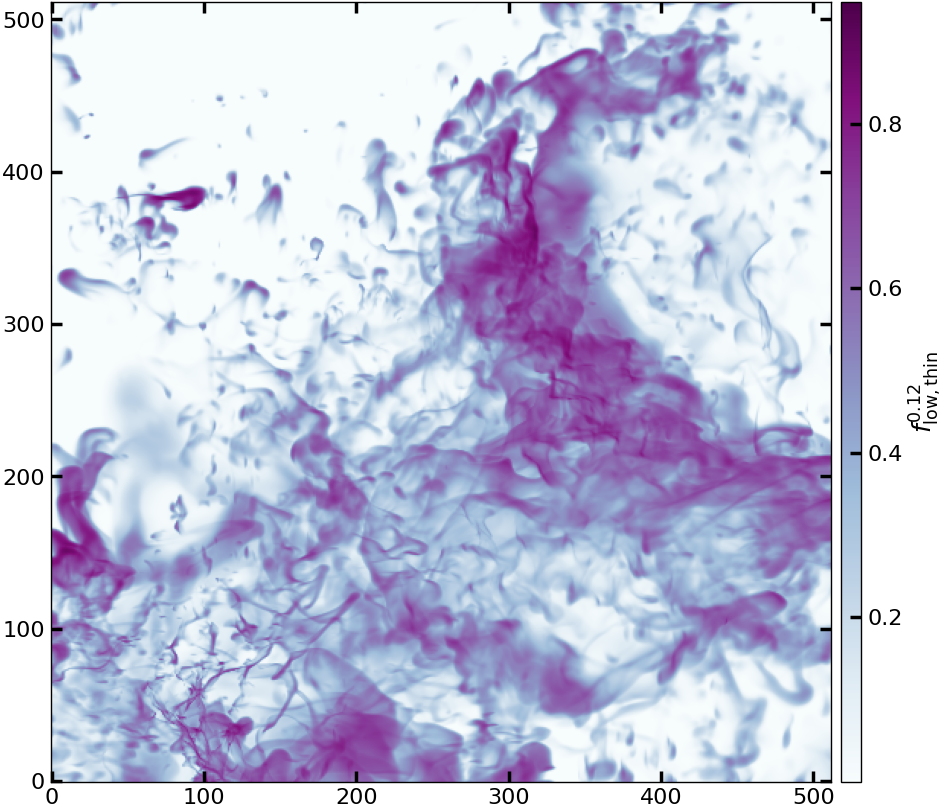} \hfill
  \includegraphics[width=0.42\linewidth]{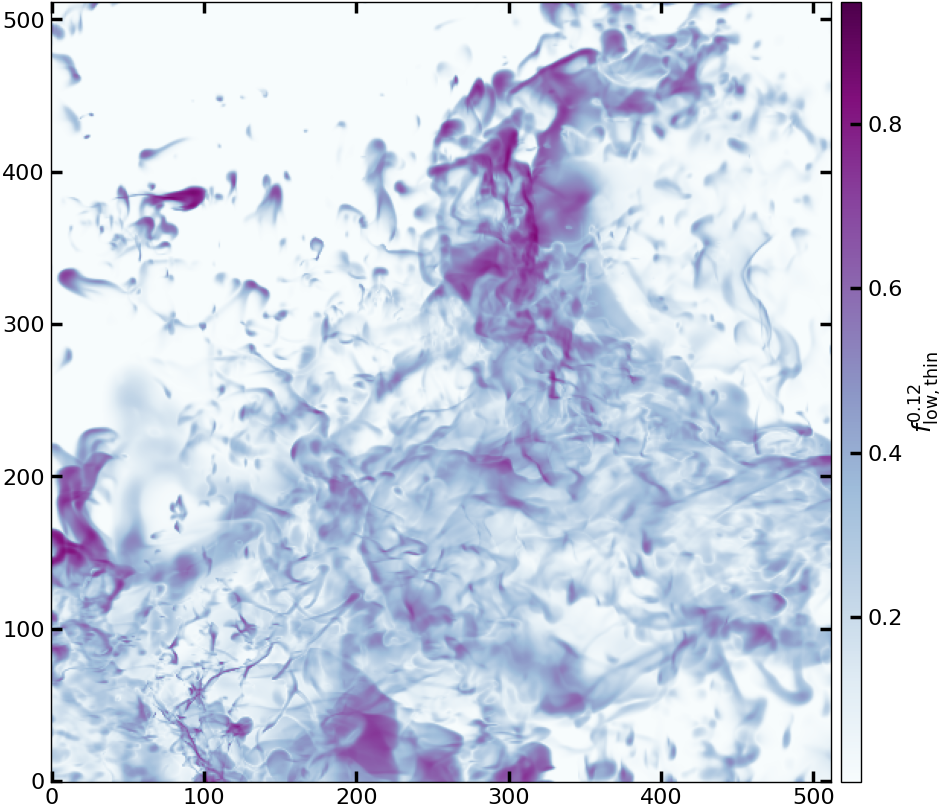}
  \includegraphics[width=0.42\linewidth]{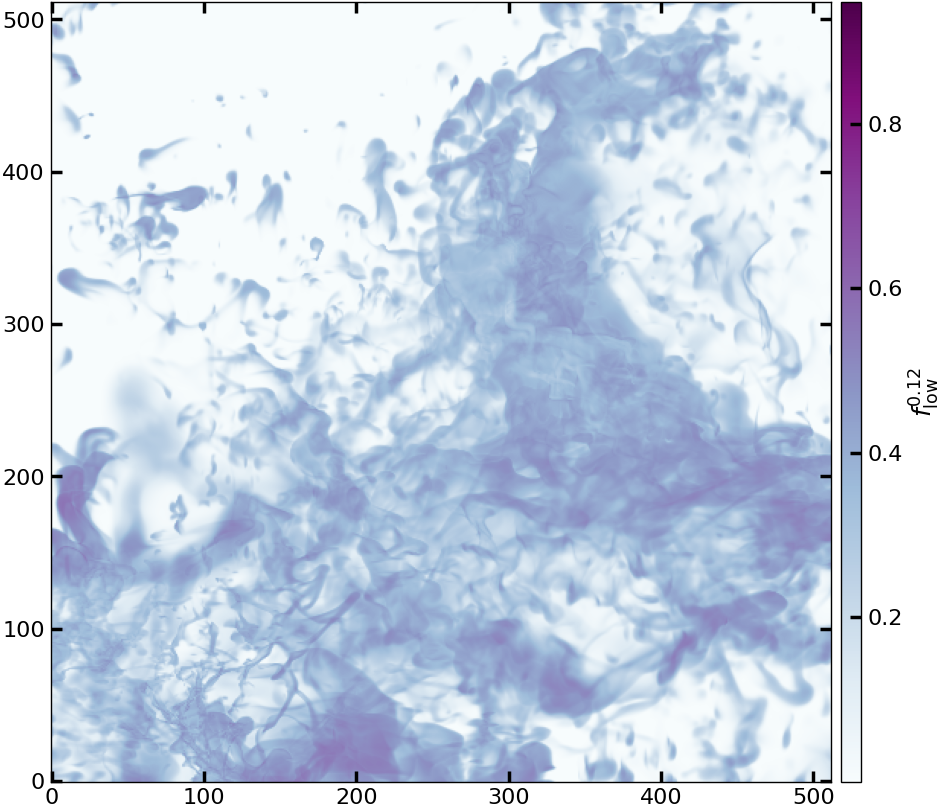} \hfill
  \includegraphics[width=0.42\linewidth]{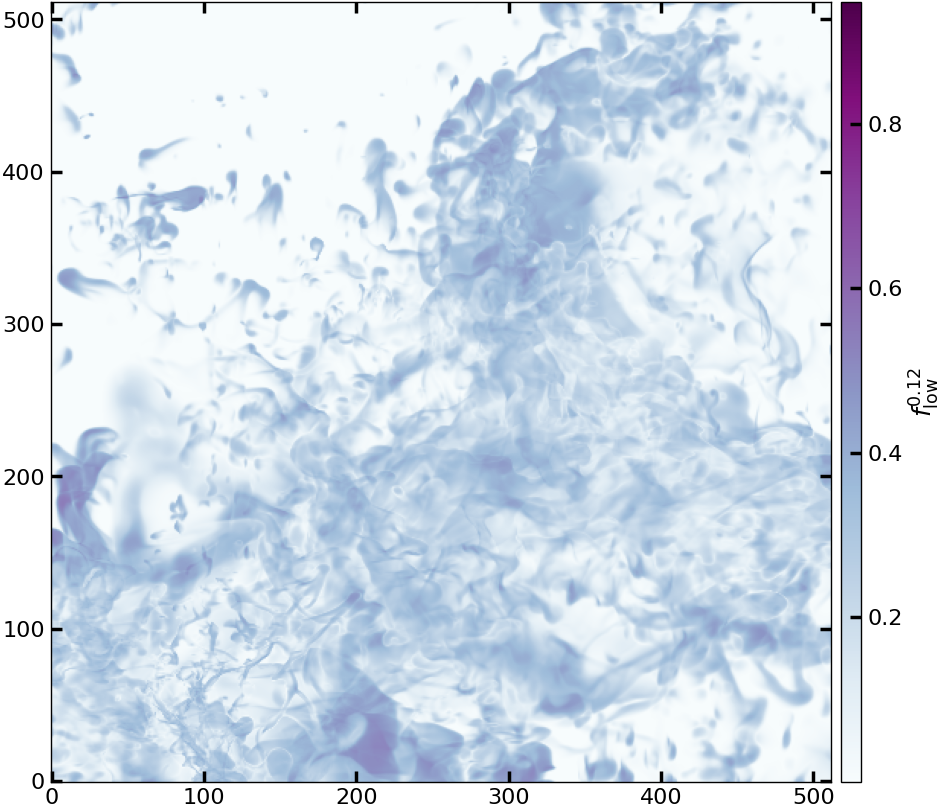}
  \caption{Map of lower limit on the cold gas mass fraction $f^{0.12}_{\rm low}$ evaluated with Equation~\ref{eq:mass_fraction} for
  $T_b^*(v_z=0,\rb)$ (top left), 
  $T_b^*(v_z,\rb)$  (top right), 
  $T_b(v_z=0,\rb)$ (bottom left), 
  and $T_b(v_z,\rb)$ (bottom right). 
  Compare to $f(T < T_{k,\rm max}$) in Figure \ref{fig:fcnm_True_saury}.
  }
  \label{fig:fcnm}
\end{figure*}

\begin{figure*}
  \centering
  \includegraphics[width=0.49\linewidth]{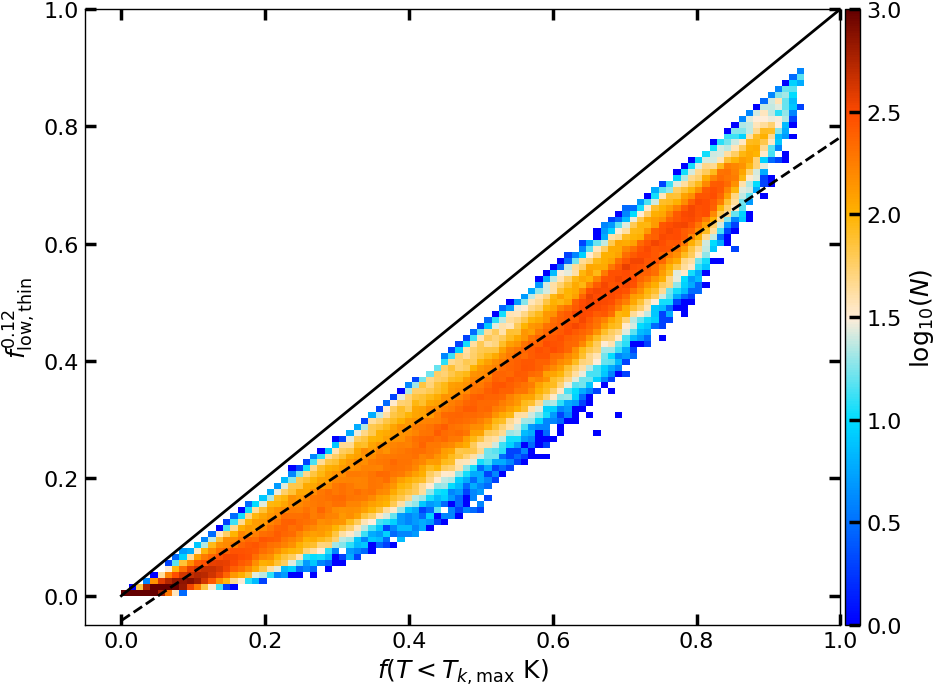}\hfill
  \includegraphics[width=0.49\linewidth]{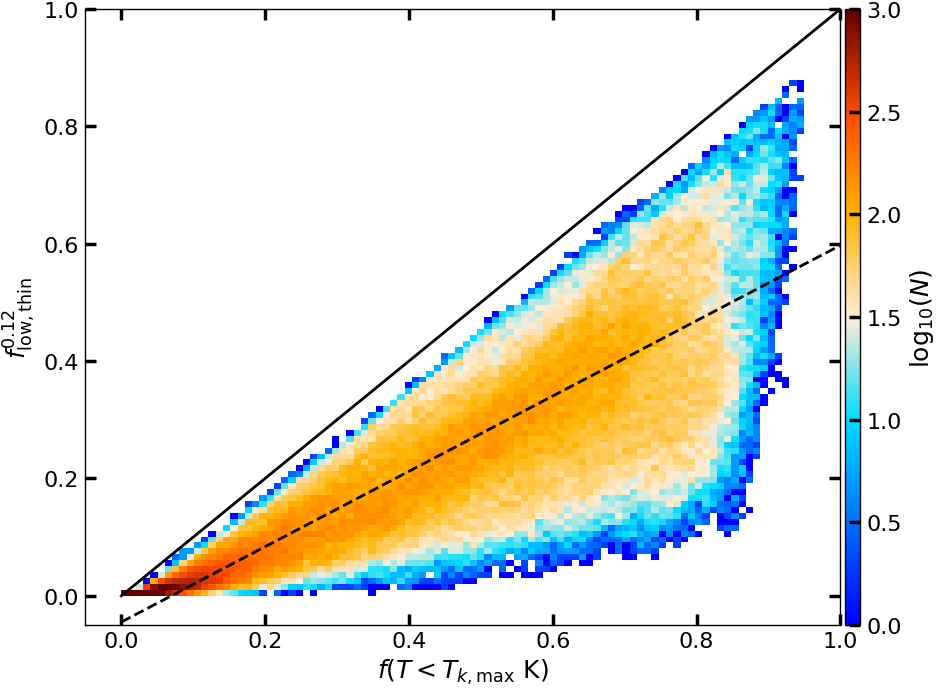}
  \includegraphics[width=0.49\linewidth]{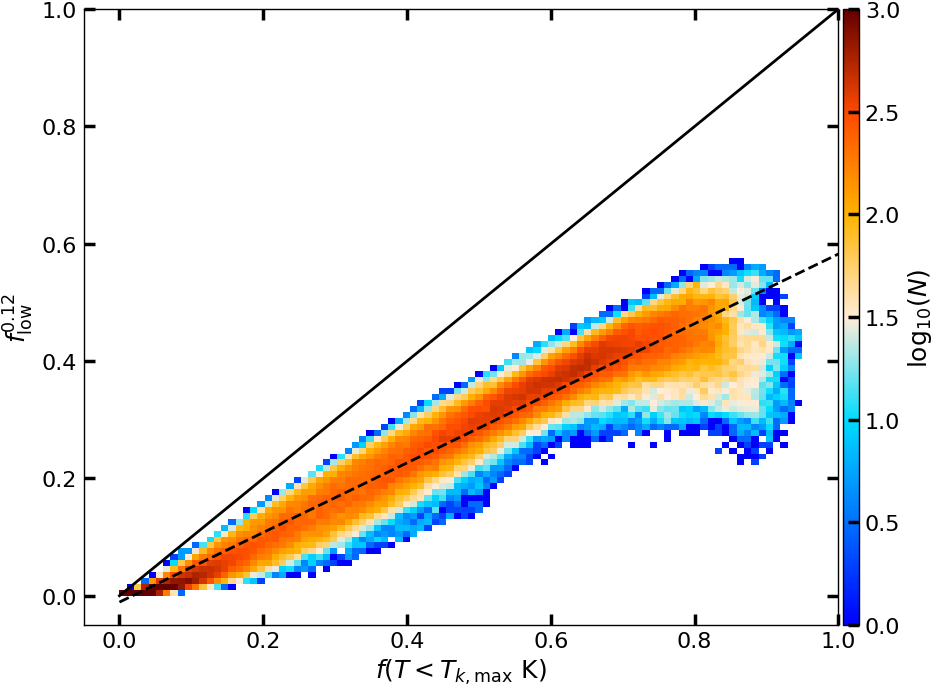}\hfill
  \includegraphics[width=0.49\linewidth]{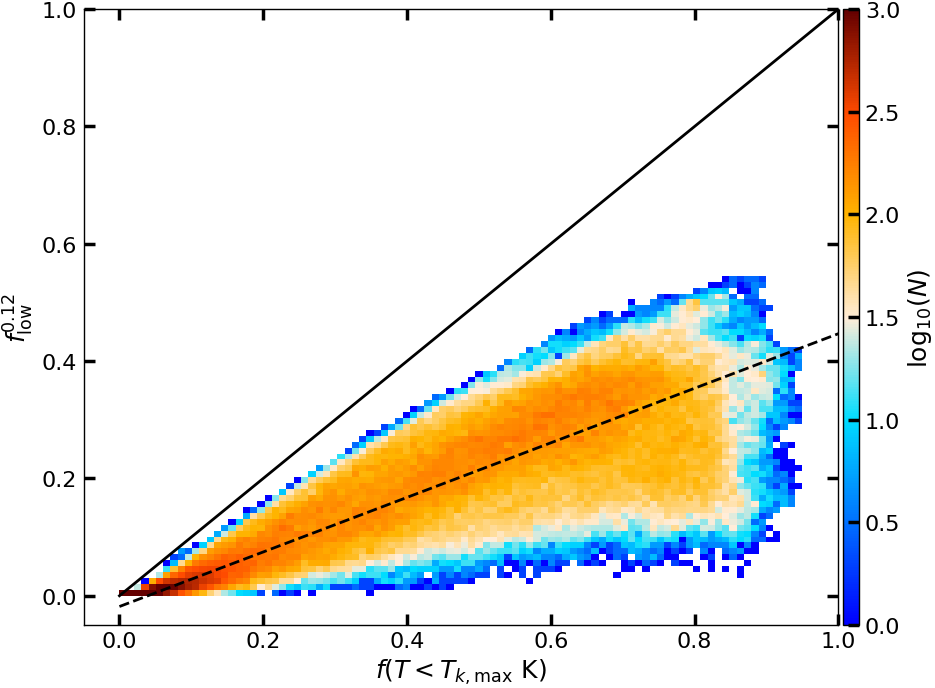}
  \caption{2D distribution function of $f^{0.12}_{\rm low}$ evaluated with Equation~\ref{eq:mass_fraction} vs.\ $f(T < T_{k,\rm max}$), for
  $T_b^*(v_z=0,\rb)$ (top left), 
  $T_b^*(v_z,\rb)$ (top right), 
  $T_b(v_z=0,\rb)$ (bottom left), 
  and $T_b(v_z,\rb)$ (bottom right).
  The solid black line shows the 1:1 line.
  Dashed lines show linear fits using the bisector estimator.
  }
  \label{fig:heatmap_CNM_fftcut}
\end{figure*}

\subsection{Testing the lower limit}
\label{subsec:simu-result}

As a baseline, Figure~\ref{fig:fcnm_True_saury} shows a projection of the mass fraction of cold gas in the simulation, obtained by integration of cells with $T < T_{k,\rm max} = 1089$\,K.  This is the ``true" cold gas mass fraction, denoted $f(T < T_{k,\rm max}$).

Figure~\ref{fig:fcnm} shows maps of $f^{0.12}_{\rm low}$ evaluated using Equation~\ref{eq:mass_fraction} on the four different synthetic spectral cubes.
For all cases, the map of the lower limit on the cold gas mass fraction strongly resembles the map of the true mass fraction obtained directly from the density field alone ($f(T < T_{k,\rm max}$)), shown in Figure~\ref{fig:fcnm_True_saury}. The spatial structure of the cold gas mass fraction is highlighted well.

This similarity is quantified in Figure~\ref{fig:heatmap_CNM_fftcut} 
which shows 2D histograms of $f^{0.12}_{\rm low}$ vs.\ $f(T < T_{k,\rm max}$) for the four different cases. Table \ref{tab:statistic} summarizes four statistics of the correlations: the Pearson correlation coefficient $r_p$, the slope and intercept, and the 
scale-averaged\footnote{The scale-averaging was performed by taking the mean of all cross correlation coefficients. Using the median (not tabulated here) yields a three percent variation around each $\left<cc(k)\right>$.} cross correlation coefficient $\left<cc(k)\right>$.
%\pgm{Add an explanation of scale-averaging, maybe in a footnote?}

Regardless of the opacity regime 
(top row or bottom row in Figures \ref{fig:fcnm} and \ref{fig:heatmap_CNM_fftcut}; columns 2 and 3 or 4 and 5 in Table \ref{tab:statistic}),
$r_p$ is the highest when $v_z(\rb,z)=0$ 
(left column in figures; columns 2 and 4 in table).
Interference patterns induced by the turbulent velocity field 
(right column in figures; columns 3 and 5 in table) lower the slope 
and spread the linear correlation with increasing ``true'' cold gas mass fraction, resulting in 
lower $r_p$. The synthetic spectra are indeed expected to be more complex (possibly showing multiple peaks) and their FTs to be more susceptible to significant interference patterns as the mass fraction of cold gas rises.

Regardless of the velocity field (left or right column), the effect of opacity (bottom row) translates into a saturation of the upper envelope to $f^{0.12}_{\rm low}$ at about 0.5.
This saturation is intrinsic to the simulated spectra where the intensity of their peaks is lowered (compared to the optically-thin treatment) by the opacity of the line. Its effect is all the more important when there is more cold gas along the line of sight.

It can be seen visually in the successive panels in Figure \ref{fig:fcnm} that despite the effects of both opacity and the velocity field, information about the projected spatial structure in Figure \ref{fig:fcnm_True_saury} is retained.
The scale-averaged cross correlation coefficients $\left<cc(k)\right>$ between $f^{0.12}_{\rm low}$ and $f(T < T_{k,\rm max}$) for the four cases are tabulated in Table~\ref{tab:statistic}. Across scales and for both treatments of opacity, the imprint of the turbulent velocity field on $f^{0.12}_{\rm low}$ removes less than 20\% of the spatial correlation with $f(T < T_{k,\rm max}$).
Qualitatively, the impact of the turbulent velocity field appears as local attenuation on $f^{0.12}_{\rm low}$. This is likely to change the statistical properties of the field (hence the statistics of the corresponding column density field).

% \pgm{I wonder if we could take the lower right panel -- all effects -- and calculate  $f(T < T_{k,\rm max} - f^{0.12}_{\rm low}$ to plot against NH. That is the diagram we want to explain later.}
% \am{See new Figure~\ref{fig:heatmap_CNM_fftcut_saury_thick_NHI_thin}. Warning: shift all Figures in referee's report. fixme if kept.}
% \pgm{Looking at this, I think I made a bad suggestion. The``true" is so much larger than the $f^{0.12}_{\rm low}$ that it is like plotting the ``true" vs.\ NH, which is not the issue at hand.  We are trying to compare the lower right panel to figure 11, other than visually.  It will never look like Figure 16. Maybe their ratio vs NH?  I think Figure 13 is enough. So let's not have the new Figure 14. Sorry.}

\section{Assessing \lowercase{$f^{0.12}_{\rm low}$} with Pointed Observations}
\label{sec:validation}

\subsection{Using $T_b$ and $\tau$ spectra from 21-SPONGE}
\label{sec:21-SPONGE-full}
%Comparison 21-SPONGE
\begin{figure}
  \centering
  \includegraphics[width=\linewidth]{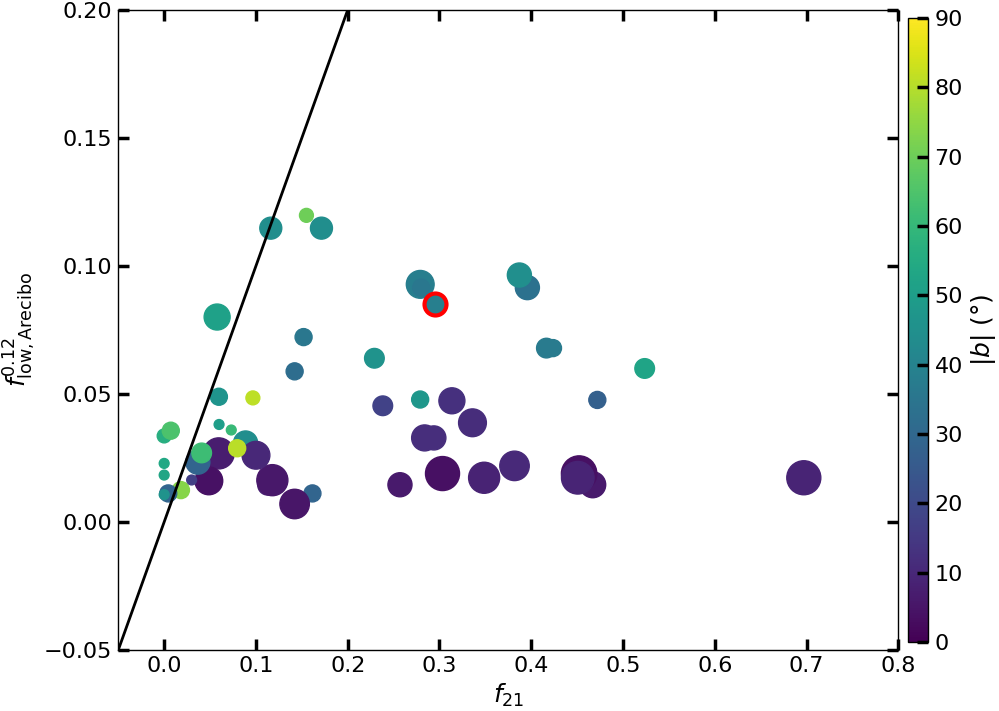}
  \caption{Scatter plot of $f^{0.12}_{\rm low, Arecibo}$ vs.\ $f_{21}$ for all sources from the 21-SPONGE survey color coded by absolute Galactic latitude.
  The size of each dot encodes the number of absorption components (ranging from 0 to 13) identified in the Gaussian decomposition of the optical depth spectrum by \citet{murray_2018}.
  The solid black line shows the 1:1 line.
  The dot with the red edges denotes the line of sight to J2232.
  }
  \label{fig:fcnm_fft_vs_f21_high_b_er}
\end{figure}

\begin{figure}
  \centering
    \includegraphics[width=\linewidth]{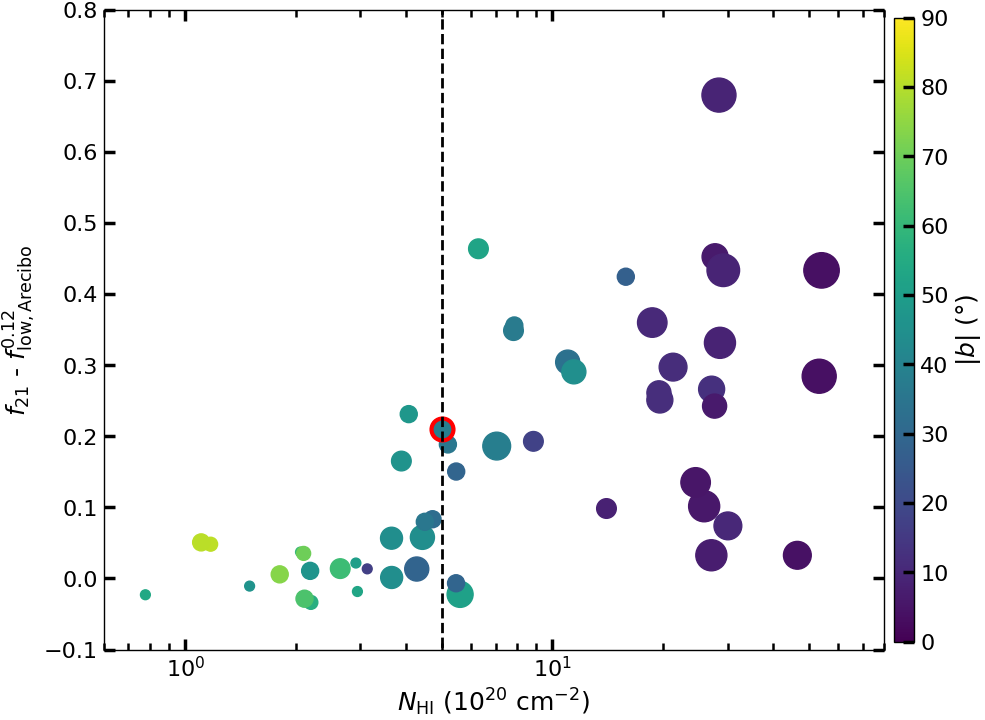}
  \caption{Scatter plot of $f_{21} - f^{0.12}_{\rm low, Arecibo}$ vs.\ $\NHI$ color coded by absolute Galactic latitude.
  The dashed vertical line denotes the typical column density where the effects of opacity become significant. 
  The size code is as in Figure~\ref{fig:fcnm_fft_vs_f21_high_b_er}.
  The dot with the red edges denotes the line of sight to J2232.
  }  
  \label{fig:fcnm_fft_vs_21sponge_high_b_er_NHI_thick_color_b}
\end{figure}

For an initial exploration of \HI\ observations, we return to the line of
sight to J2232 (Galactic latitude $-38\fdg582$) discussed in Section \ref{subsec:21-sponge}, whose 
21-SPONGE survey $\tau$ and interpolated $T_b$ spectra and FTs are displayed in Figures \ref{fig:diffraction_21Sponge_Tb} and \ref{fig:diffraction_21Sponge}. 
\citet{murray_2018} carried out a fairly classical joint analysis of $\tau$ and $T_b$ spectra, including Gaussian decomposition and exploiting the fact that the optical depth is inversely dependent on the spin temperature $T_s$, to obtain a cold gas mass fraction of $f_{21} = 0.30\pm0.10$.
%above our lower limit.
For our simple estimate, using only the FT of the interpolated emission spectrum $T_b$, Equation~\ref{eq:mass_fraction} for the lower limit evaluates to $f^{0.12}_{\rm low, Arecibo}=0.08$, indeed lower than $f_{21}$, which is expected to be closer to the true value.

The 21-SPONGE dataset covers lines of sight over a range of Galactic latitude with spectra displaying a variety of complexity.
We extended the comparison made for J2232 to the entire dataset.
A scatter plot is shown in Figure~\ref{fig:fcnm_fft_vs_f21_high_b_er}, 
where dots are color coded by absolute Galactic latitude and their sizes encode the complexity of the $\tau$ spectra (the number of absorption components, ranging from 0 to 13, identified in the Gaussian decomposition by \citealt{murray_2018}).
The dot with the red edges denotes the line of sight to J2232.

Most of the $f^{0.12}_{\rm low, Arecibo}$ values are below the 1:1 line shown by the solid black line (i.e., $<f_{21}$), which supports our proposition that $f^{0.12}_{\rm low}$ would provide a lower limit.
The lower limit can approach the true value in some circumstances: at high Galactic latitudes and relatively low cold gas mass fraction ($f_{21}\lesssim0.2$) and spectral complexity (see encoding of dots), we observe a fair agreement between $f^{0.12}_{\rm low, Arecibo}$ and $f_{21}$. 
See also the discussion of Figure~\ref{fig:fcnm_fft_vs_21sponge_high_b_er_NHI_thick_color_b} below. However, for $f_{21} \gtrsim 0.2$ and increasing spectral complexity, $f^{0.12}_{\rm low, Arecibo}$ saturates below about 0.1.
At lower Galactic latitudes the trend is similar, but the values of  $f^{0.12}_{\rm low, Arecibo}$ are systematically lower for a given $f_{21}$ and saturate below about 0.03.
This relative behavior of $f^{0.12}_{\rm low, Arecibo}$ is plausibly due to a combination of increasing opacity coupled with a higher complexity of the emission line (i.e., more components seen in absorption, as shown by the size of each dot, is likely associated with a more complex line profile in emission as well), leading to more interference in the FT.

In the Taurus and Gemini Regions, \cite{nguyen_2019} observed a strong dependency of the correction to the optically thin \HI\ approximation, $\mathcal R=\NHI/\NHI^*$, on the cold gas mass fraction. The authors found that for the CNM mass fraction below 0.2, $\NHI^*$ and $\NHI$ are consistent within the errors. However, for mass fraction higher than 0.2 (and lower than 0.75), $\mathcal{R}$ rises from 1 to 2.

A scatter plot of the difference $f_{21} - f^{0.12}_{\rm low, Arecibo}$ vs.\ total column density $\NHI$ so corrected for opacity is shown in Figure~\ref{fig:fcnm_fft_vs_21sponge_high_b_er_NHI_thick_color_b}.
Values of $f_{21}$ and $f^{0.12}_{\rm low, Arecibo}$ are fairly similar for $\NHI\lesssim 3-5\times10^{20}$\,cm$^{-2}$, which is close to the typical column density, $\NHI\approx5\times10^{20}$\,cm$^{-2}$, where opacity effects start to become apparent in various absorption line surveys \citep[see e.g., figure 2 in][showing $\mathcal{R}$ vs.\ the opacity-corrected $\NHI$ for data in the Arecibo Millennium Survey \citep{heiles_millennium_2003} and 21-SPONGE]{nguyen_2018}.
That means that at low column density either our lower limit is close to the true value, which it would tend to be for narrow lines from cold gas such as are favored by absorption line measurements (also low turbulent broadening), or the true value $f_{21}$ is also an underestimate.
The dot with the red edges denotes the line of sight to J2232 (Figure \ref{fig:diffraction_21Sponge}), near this threshold column density.
The effect of the combination of increasing opacity coupled with a higher complexity of the line (dot size), especially at low latitude, can be appreciated here again.\footnote{We observe a similar trend (not shown here) when plotted instead against the total column density $\NHI^*$ computed in the optically thin limit.}

Although values of $f_{21}$ and $f^{0.12}_{\rm low, Arecibo}$ are fairly similar at low column density, we do note that seven lines of sight have $f_{21} - f^{0.12}_{\rm low, Arecibo} < 0$, which seems to suggest an inconsistency between our lower limit and the cold gas mass fraction measured in absorption.
We compared values of this difference with the uncertainties on the cold gas mass fraction calculated by \citet[][see their table D1]{murray_2020}. Four of them are consistent within errors but three lines of sight remain unexplained, namely 3C245A, 1055+018, and 3C273.
We have explored this further in Section~\ref{sec:validation-bighicat} (validation on BIGHICAT) where a significant number of lines of sight show the same behavior and cannot be reconciled within uncertainties.

\subsection{Lower resolution emission line data from HI4PI}
\label{sec:validation-hi4pi}

We evaluated the applicability of our FT method to 21\,cm emission line data with lower spatial and spectral resolution than 21-SPONGE (3\farcm5), namely that from the HI4PI survey,\footnote{HI4PI is a combined all-sky product based on data from the Effelsberg-Bonn \HI\ Survey \citep[EBHIS, surveyed with the Effelsberg 100-meter radio telescope;][]{kerp_2011,winkel_effelsberg-bonn_2016} and from the third revision of the Galactic All-Sky Survey \citep[GASS, surveyed with the Parkes telescope;][]{mcclure-griffiths_2009,kalberla_2010_gass,kalberla_haud_2015_gass}.} which has a 16\farcm2 beam and 1.49\kms\ spectral resolution \citep{hi4pi_2016}.
At the position of each source from 21-SPONGE for which we previously evaluated $f^{0.12}_{\rm low, Arecibo}$, we extracted the on-source $T_b$ spectrum in the spectral range $-90 < v < 90$\,\kms, apodizing it with a cosine function at both ends (Section \ref{subsec:apod}).
On this extracted HI4PI spectrum, we evaluated our FT based estimate of the lower limit of the cold gas mass fraction, $f^{0.12}_{\rm low, HI4PI}$.

\begin{figure}
  \centering
    \includegraphics[width=\linewidth]{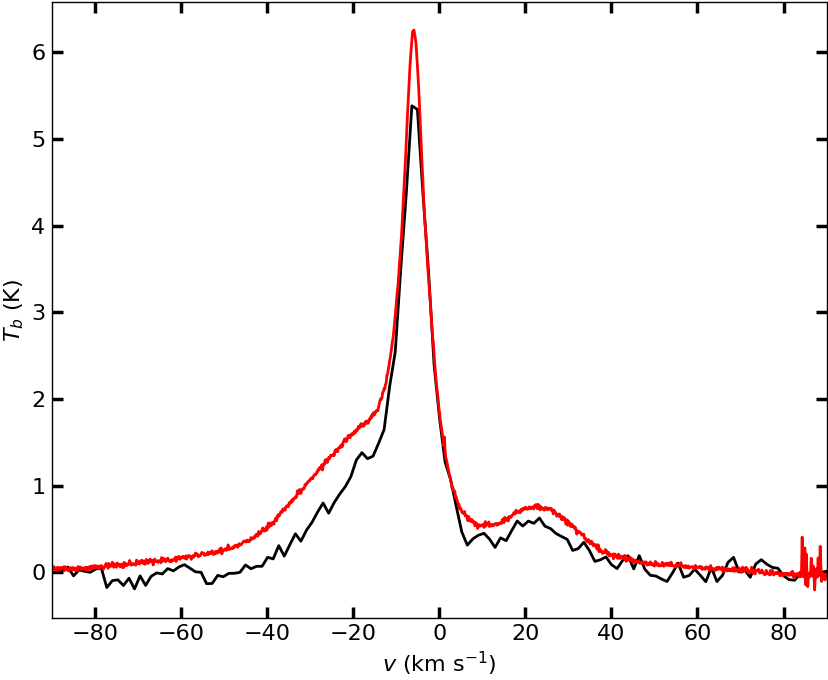}
    \caption{
    Comparison between the brightness temperature spectrum interpolated around 3C273 (source with the highest $T_c$) from 21-SPONGE (red) and the on-source spectrum extracted from the HI4PI survey (black).
    }
  \label{fig:Tb_onoff}
\end{figure}

Using the on-source spectrum rather than an interpolated spectrum assumes that at the low spatial resolution of HI4PI the peak brightness temperature of the continuum source $T_c$ is diluted so much that the on-source emission spectra do not show the effects of absorption.
To quantify this, $T_c$ seen by HI4PI at 16\farcm2 is \citep{blagrave_dhigls:_2017} 
\begin{equation}
    T_c = 170 \, \left( \frac{[1']}{16\farcm2} \right)^2 \, \frac{S_\nu}{[1\,\rm{Jy}]} \, \rm{K} \,,
    \label{eq:Tc}
\end{equation}
where $S_\nu$ is the source flux density.
The background source is reduced by \HI\ absorption producing a negative-going feature in the continuum-subtracted spectrum with profile $T_c\,(1 - \exp(-\tau))$.  This is to be compared to the emission line profile, which in a gas of constant spin temperature $T_s$ is $T_b = T_s\,(1 - \exp(-\tau))$.  Therefore, we want to check that $T_c$ is significantly less than $T_s$.  We adopt $T_s = 88$\,K for cold gas \citep{blagrave_dhigls:_2017} and note that even for the
brightest source in 21-SPONGE, 3C273, $T_c = 36$\,K.

As an empirical check, in Figure~\ref{fig:Tb_onoff} we have compared the brightness temperature spectrum interpolated from off-source spectra around 3C273 from 21-SPONGE (red) and the on-source spectrum extracted from the HI4PI survey (black).
Their peak brightness temperatures differ by less than 1\,K.
Importantly for our estimator of the cold gas mass fraction, the value $f^{0.12}_{\rm low, HI4PI} = 0.06$ for the black spectrum is close to $f^{0.12}_{\rm low, Arecibo} = 0.036$ for the red spectrum.

In addition, we extracted a 2.56 deg$^2$ HI4PI brightness temperature PPV cube (not shown here) centered on the sources and found no evidence of attenuation in individual channel maps.

\begin{figure}
  \centering
    \includegraphics[width=\linewidth]{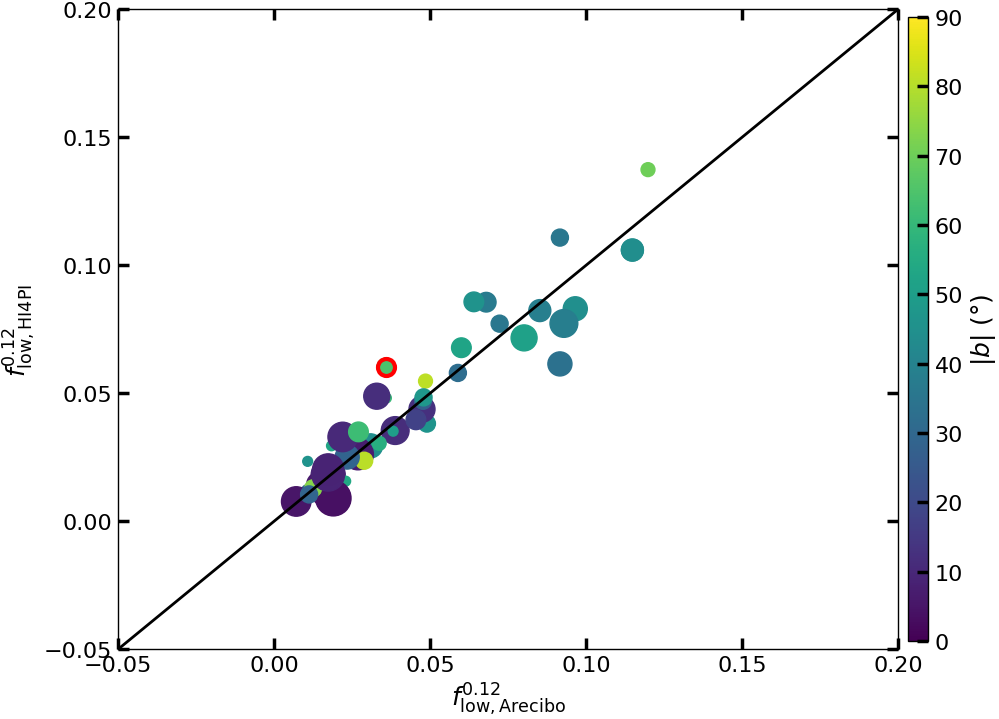}
    \caption{Scatter plot of $f^{0.12}_{\rm low, HI4PI}$ vs.\ $f^{0.12}_{\rm low, Arecibo}$ for lines of sight in 21-SPONGE.
    Color code, size, and annotations are as in Figure~\ref{fig:fcnm_fft_vs_f21_high_b_er}.
    The dot with the red edges denotes the line of sight to 3C273  shown in Figure~\ref{fig:Tb_onoff}.
 }
  \label{fig:fcnm_fft_HI4PI_vs_Arecibo}
\end{figure}

In summary, Figure~\ref{fig:fcnm_fft_HI4PI_vs_Arecibo} plots $f^{0.12}_{\rm low, HI4PI}$ vs.\ $f^{0.12}_{\rm low, Arecibo}$. Points are scattered around the 1:1 relation shown by the black solid line with a relatively low dispersion that shows that $f^{0.12}_{\rm low}$ is quite consistent over spatial resolutions varying from 3\farcm5 to 16\farcm2, at least for lines of sight in 21-SPONGE.
Although the cold gas emission is likely to arise in small scale structures, its fractional contribution is not being modified significantly within a 16\farcm2 beam. Furthermore, despite the lower velocity resolution of HI4PI, the cold gas with narrow lines is being reliably detected.

\subsection{Using a CNN on $T_b$ spectra} %Comparison with Murray et al. (2020)}
\label{sec:CNN}
\begin{figure}
  \centering
    \includegraphics[width=\linewidth]{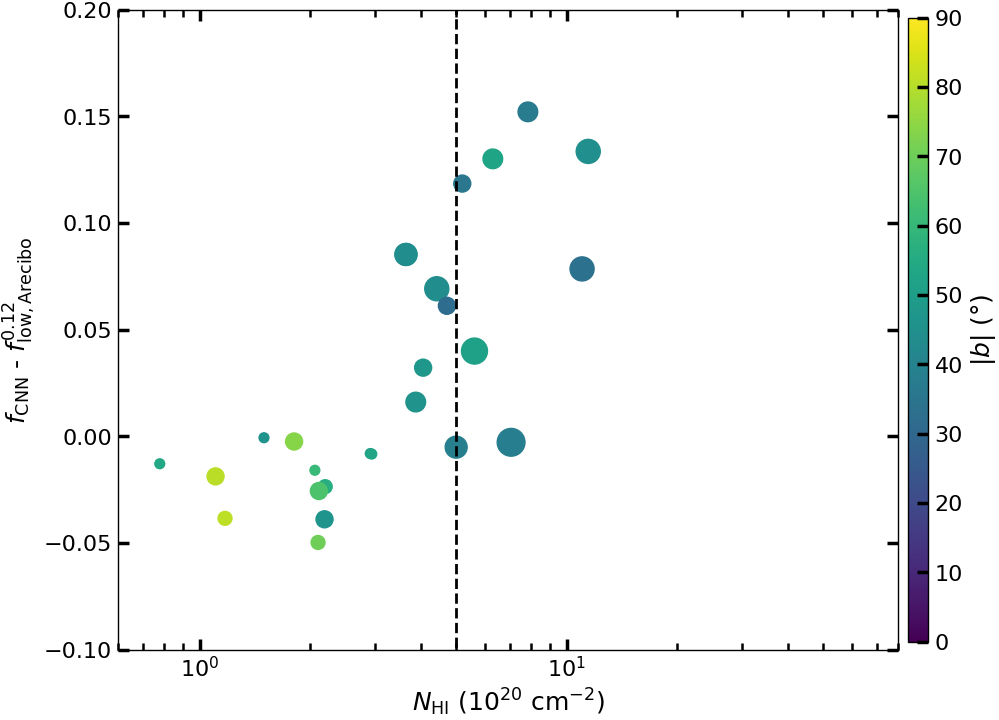}
    \caption{Scatter plot of $f_{\rm CNN} - f^{0.12}_{\rm low, Arecibo}$ vs.\ $\NHI$.
    Color code, size, and annotations are as in Figure~\ref{fig:fcnm_fft_vs_21sponge_high_b_er_NHI_thick_color_b}.
%\pgm{Limit upper y range to 0.2.}
}  
    \label{fig:fcnm_fft_vs_CNN_high_b_er_NHI_thick}
\end{figure}

As mentioned in the introduction, \cite{murray_2020} developed a 1D CNN, trained on simulated data that included opacity information and designed to infer an absolute value of the cold gas mass fraction when applied to emission line data.
For the same 21-SPONGE emission data for the J2232 line of sight, using this 1D CNN they 
obtained $f_{\rm CNN}=0.08\pm0.03$, similar to the lower limit obtained with our FT method.
Both $f_{\rm CNN}$ and $f^{0.12}_{\rm low, Arecibo}$ are lower than $f_{21}$.

We used table D1 in \cite{murray_2020} to produce the 
scatter plot of $f_{\rm CNN} - f^{0.12}_{\rm low, Arecibo}$ against $\NHI$ shown in Figure~\ref{fig:fcnm_fft_vs_CNN_high_b_er_NHI_thick} 
for all the lines of sight at high Galactic latitude (i.e., $|b|>30$\,\degree) used in \cite{murray_2020}.
The color code, size, and annotations are as in Figure~\ref{fig:fcnm_fft_vs_21sponge_high_b_er_NHI_thick_color_b}.
Both the CNN and our FT method are consistent within 10\% for $\NHI\lesssim 3-5\times10^{20}$\,cm$^{-2}$. The FT values are lower than the CNN values above this threshold, which again supports our proposition that $f^{0.12}_{\rm low}$ is a lower limit, affected by both the opacity of the line and the complexity of the line shape.

\subsection{BIGHICAT}
\label{sec:validation-bighicat}

\begin{figure}
  \centering
  \includegraphics[width=\linewidth]{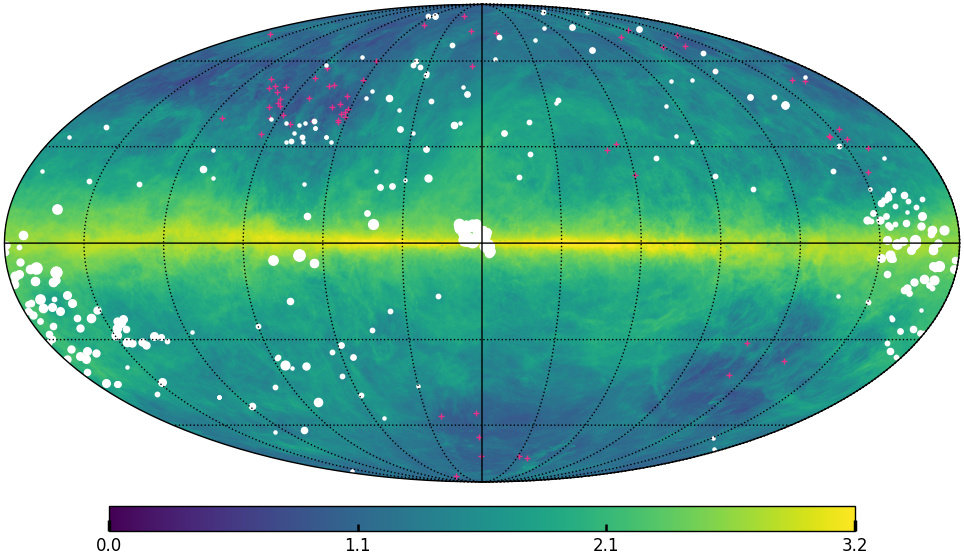}
  \caption{Mollweide projection centered on the Galactic center of the total column density map $\NHI^*$ (in units of 10$^{19}$\,cm$^{-2}$, on a logarithmic scale) from the HI4PI survey over the velocity range $-90 < v < 90$\,\kms.
  The purple crosses show the positions of non-detections in the compilation of \HI\ absorption spectra from the BIGHICAT meta-catalog \citep{mcclur23}. 
  The white dots, whose size encodes the number of absorption components (ranging from 1 to 16) in their $\tau$ spectrum, show the detections in BIGHICAT.}
\label{fig:NHI_TOT_LOWER_HI4PI_mollview}
\end{figure}

\begin{figure}
  \centering
    \includegraphics[width=\linewidth]{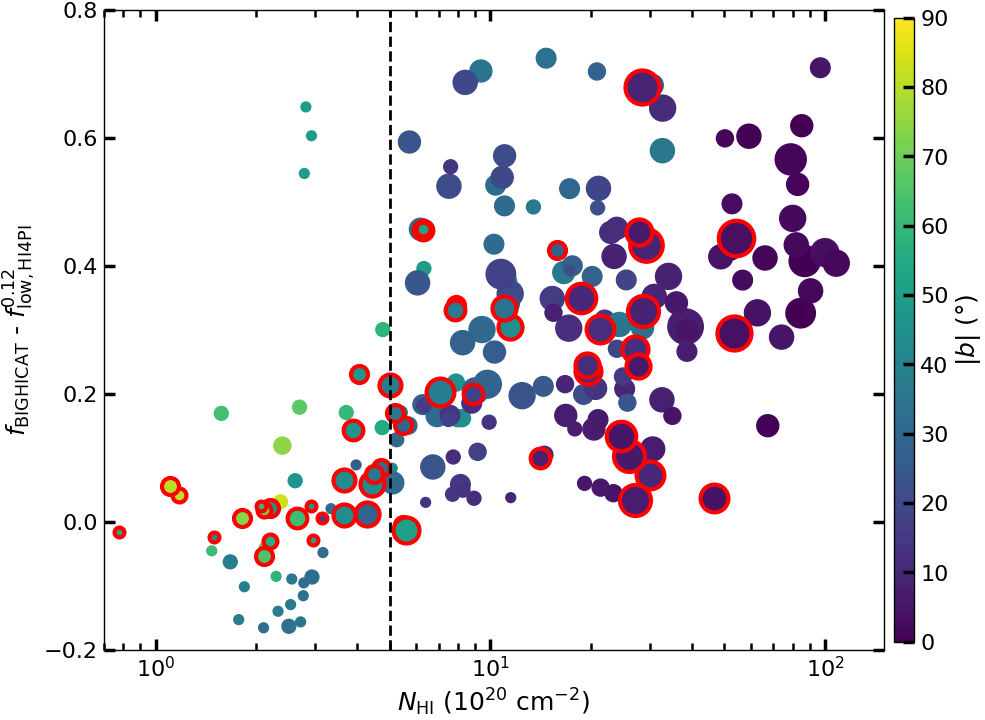}
    \caption{Scatter plot of $f_{\rm BIGHICAT} - f^{0.12}_{\rm low, HI4PI}$ vs.\ $\NHI$ color coded by absolute latitude.
    Color code, size, and annotations are as in Figure~\ref{fig:fcnm_fft_vs_21sponge_high_b_er_NHI_thick_color_b}.
    The dots with red edges show lines of sight from 21-SPONGE.
    The locations of lines of sight with a negative difference are shown with green dots in Figure~\ref{fig:f_CNM_LOWER_HI4PI_mollview}.
    }  
    \label{fig:fcnm_fft_vs_BIGHICAT_high_b_er_NHI_thick}
\end{figure}

We extended the comparison made with 21-SPONGE to the 373 unique lines of sight in the BIGHICAT meta-catalog compiled by \citet{mcclur23}.

%BIGHICAT
BIGHICAT combines results from absorption surveys with publicly available spectral Gaussian modeling \citep{heiles_millennium_2003,mohan_2004,stanimirovic_2014,denes_2018,murray_2018,nguyen_2019,murray_2021}.
As \citet{mcclur23} emphasize, BIGHICAT is heterogeneous in terms of sensitivity 
with median RMS noise in \HI\ optical depth per 1\kms\ channel from 0.0006 (21-SPONGE) to 0.1 \citep{denes_2018}, angular resolution from $1^\prime$ \citep[CGPS and VGPS, ][]{taylor_2003,stil_2006} to $36^\prime$  \citep{roy_temperature_2013,mohan_2004a,mohan_2004b}, velocity resolution from 0.1\kms\ \citep{denes_2018} to 3.3\kms\ \citep{mohan_2004a,mohan_2004b}, and the Gaussian decomposition method used to model the data (all but \citealt{mohan_2004} and \citealt{denes_2018} follow the methodology developed by \citealt{heiles_millennium_2003}).
While this might affect the homogeneity of the cold gas mass fraction estimates, $f_{\rm BIGHICAT}$, it provides a larger sample that probes different regions of the sky.

Figure~\ref{fig:NHI_TOT_LOWER_HI4PI_mollview} shows the location of the 373 lines of sight in BIGHICAT on a Mollweide projection of the total column density map $\NHI^*$ (on a logarithmic scale) from the HI4PI survey over the velocity range $-90 < v < 90$\,\kms.
The purple crosses show the positions of non-detections and the white dots, whose sizes encode the number of absorption components (ranging from 1 to 16) in their $\tau$ spectrum, show the detections. 

For comparison to $f_{\rm BIGHICAT}$, we need an emission spectrum on which to apply our FT method. For a homogeneous product, we used on-source spectra from HI4PI as described in Section \ref{sec:validation-hi4pi}.
Figure~\ref{fig:fcnm_fft_vs_BIGHICAT_high_b_er_NHI_thick} shows the scatter plot of $f_{\rm BIGHICAT} - f^{0.12}_{\rm low, HI4PI}$ vs.\ $\NHI$ color coded by absolute latitude for all lines of sight in BIGHICAT. Dots with red edges show lines of sight from 21-SPONGE. Due to the high correlation observed between $f^{0.12}_{\rm low, HI4PI}$ and $f^{0.12}_{\rm low, Arecibo}$ (see Figure~\ref{fig:fcnm_fft_HI4PI_vs_Arecibo}), the distribution for 21-SPONGE lines of sight is similar to that in Figure~\ref{fig:fcnm_fft_vs_21sponge_high_b_er_NHI_thick_color_b}.
The comparison with BIGHICAT as a whole shows a similar trend to that of 21-SPONGE only, but many more lines of sight.
Again our explanation of the scatter at large column density is that $f^{0.12}_{\rm low, HI4PI}  < f_{\rm BIGHICAT}$, with the amount of underestimation by our lower limit $f^{0.12}_{\rm low, HI4PI}$ depending on the opacity and the destructive interference introduced by the complexity of the line profile.

However, at low column density there are now many sight lines where our lower limit is higher than the cold gas mass fraction tabulated in BIGHICAT, by nearly 20\% (as opposed to less than 10\% for 21-SPONGE only). 
On investigating their distribution on the sky, we found most to be part of the MACH survey \citep{murray_2021} that targets a small region of the sky located in the first quadrant above the Galactic plane (see the concentration of purple crosses and white dots in Figure~\ref{fig:NHI_TOT_LOWER_HI4PI_mollview}). 
This is a region where we found a large cloud of relatively low column density and high cold gas mass fraction when evaluating Equation~\ref{eq:mass_fraction} on HI4PI over the whole sky (see Section~\ref{sec:hi4pi} and Figure~\ref{fig:f_CNM_LOWER_HI4PI_mollview} therein).

Interestingly, in their analysis  whose results are incorporated into BIGHICAT \citet{murray_2021} reported a case (source named J14434) where their Gaussian modeling of the spectra yielded no detection of CNM in absorption at local velocities while the emission data from EBHIS at 10\farcm8 resolution show a clear narrow peak at the same velocities. This can be appreciated in their figure~2 (first column and second row). 
They suggested that this is likely to reflect the nature of small scale structure of the CNM in this region, i.e., a porous medium in which the pencil beam of the absorption line observation misses the CNM structure in the large beam.\footnote{Depending on the spatial structure, this seemingly paradoxical situation might be avoided by an emission-line survey at very high spatial resolution. However, in that case $T_c$ of the radio source would be very high and any absorption would compromise the on-source spectrum. Interpolating emission spectra from the surrounding annulus would reintroduce the issue of sampling different spatial structures.}
As a result, $f_{\rm BIGHICAT} = 0$ at local velocities and 0.01$\pm$0.02 when considering the full velocity range (incl. an IVC component at about $-40$\,\kms).
The case of J14434 suggests that other lines of sight in this region are likely to be affected by the same effect.% and therefore explain why they have a lower limit that is higher than the cold gas mass fraction tabulated in BIGHICAT.}

EBHIS data appear in HI4PI at degraded 16\farcm2 resolution and it seems reasonable that a 16\farcm2 beam probing the same porous medium would exhibit the same effect in emission.
The HI4PI spectrum at the position of the source (not shown here) is similar to that of the EBHIS data, also showing a narrow peak at local velocities.
In that specific case, our FT approach finds $f^{0.12}_{\rm low, HI4PI} = 0.1$.
Because of the underlying structure of the phases within the 16\farcm2 beam, $f^{0.12}_{\rm low, HI4PI}  >  f_{\rm BIGHICAT}$ but $f^{0.12}_{\rm low, HI4PI}$ is still a lower limit at the beam size at which the FT was applied.

\section{Assessing \lowercase{$f^{0.12}_{\rm low}$ using the \ROHSA\ Thermal Phase Decomposition of LLIV1}}
\label{sec:validationrohsa}

\begin{figure}
  \centering
  \includegraphics[width=\linewidth]{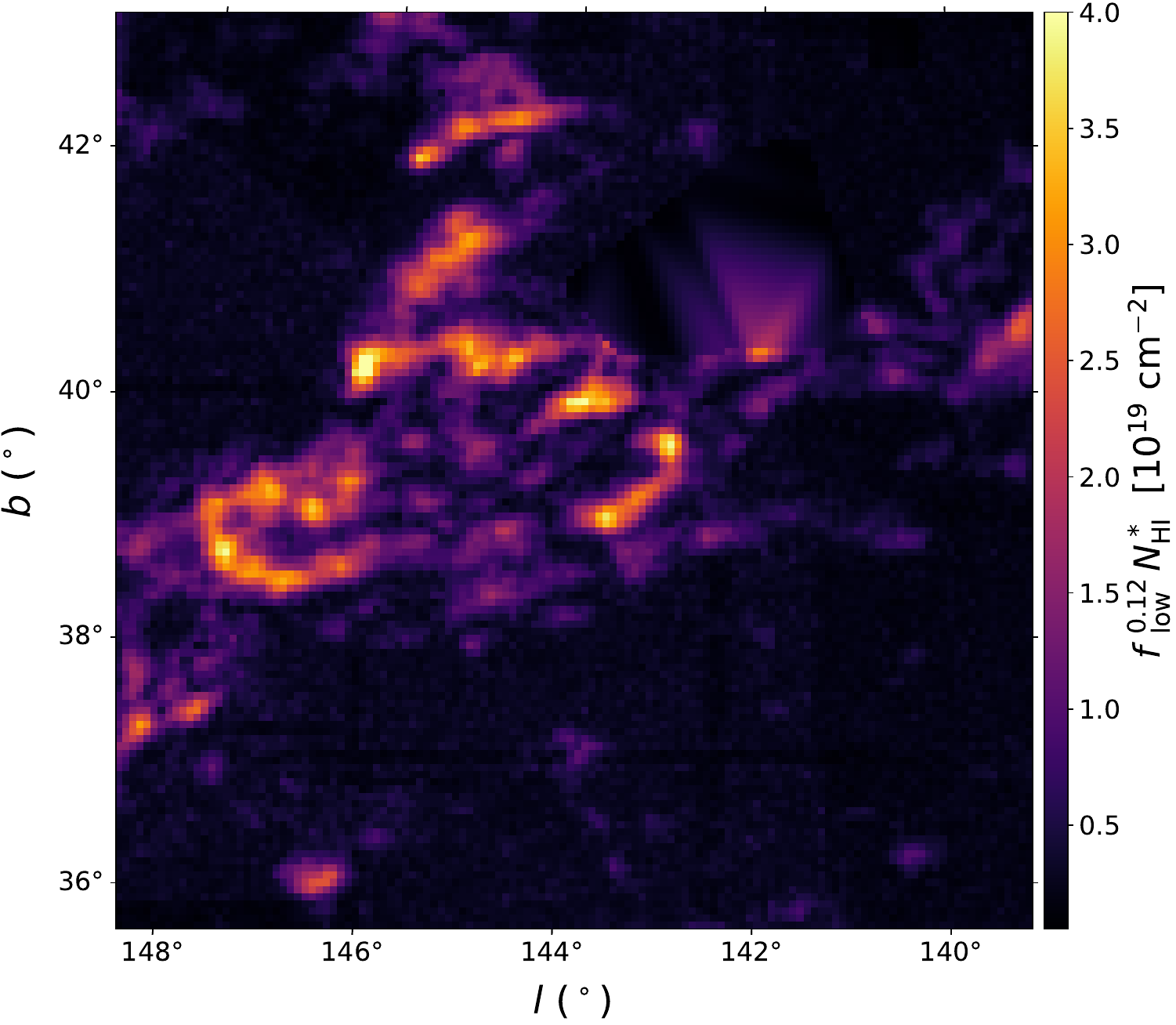}
  \caption{Lower limit on the cold gas column density $f^{0.12}_{\rm low}\,\NHI^*$ of Low-Latitude Intermediate-Velocity Arch 1 from GHIGLS in the velocity range $-81 < v < -27$\,\kms.  The color bar is in units $10^{19}$\,cm$^{-2}$.
  }
  \label{fig:LLIV1_CNM}
\end{figure}

\citet{vujeva_2023} used the spectral decomposition code \ROHSA\ (see details in Section \ref{subsec:denoising}) to
model the column density of different thermal phases in the Low-Latitude Intermediate-Velocity Arch 1 (LLIV1) dataset from GHIGLS in the velocity range $-81 < v < -27$\,\kms\ \citep{vujeva_2023}.
At each pixel, the \ROHSA\ decomposition is expressed as Gaussian components, each with an amplitude, central velocity ($\mu$), and velocity dispersion ($\sigma$).  These parameters cluster in a 2D histogram of $\sigma$ vs.\ $\mu$ (the $\sigma - \mu$ diagram, figure 6 in \citealp{vujeva_2023}). The Gaussians in clusters with $\sigma \sim 2$ \kms\ are interpreted as from CNM gas, leading to  
a map of the column density of CNM gas, their figure 7 (upper left); that map,  $N^*_{\mathrm{HI_{CNM,\, ROHSA}}}$, is the basis for the test here.  
We have applied the FT method to the same data cube and produced a map of $f^{0.12}_{\rm low}\,\NHI^*$, shown in Figure \ref{fig:LLIV1_CNM}.

\begin{figure}
  \centering
  \includegraphics[width=\linewidth]{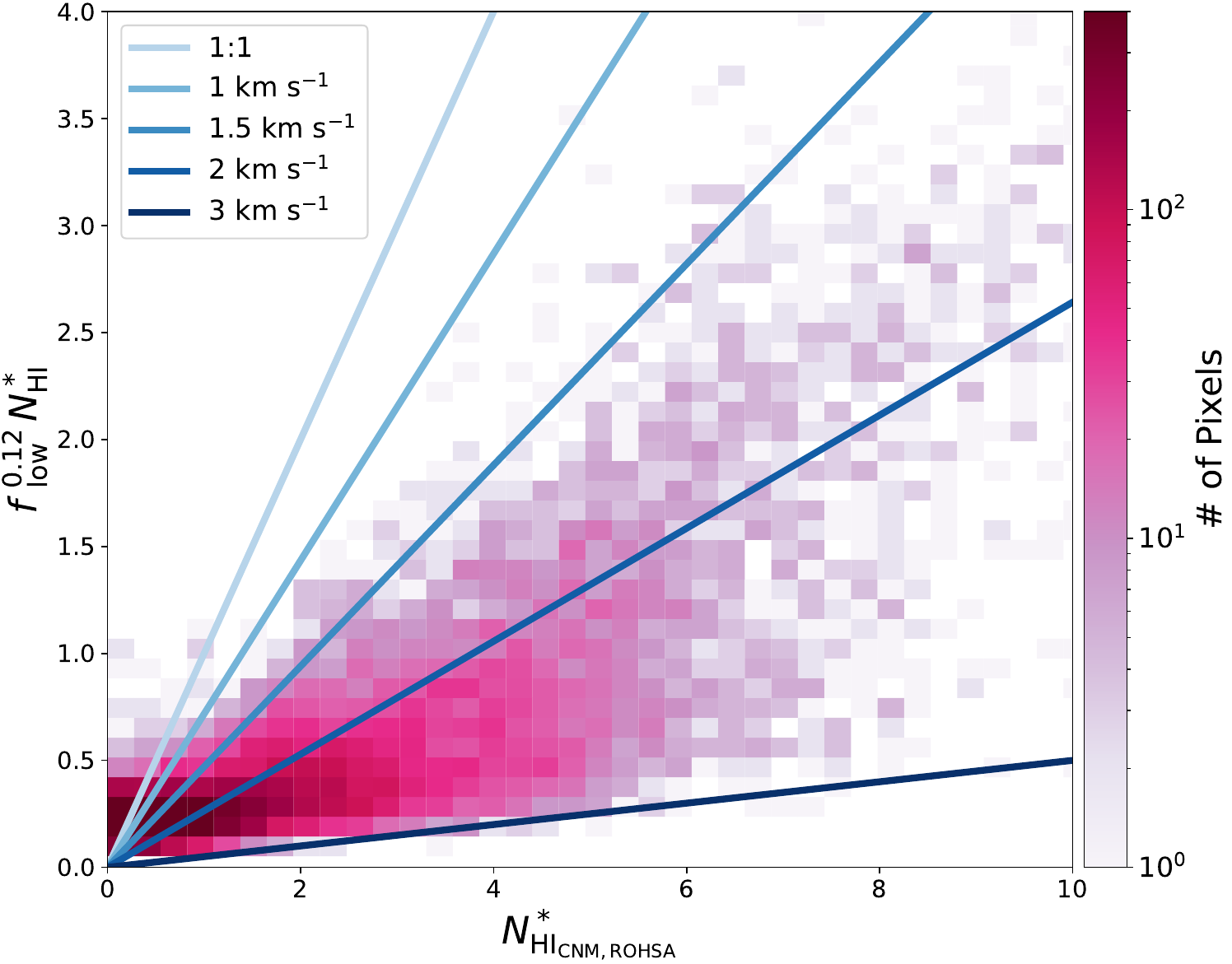}
  \caption{2D histogram (logarithmic color bar) demonstrating the pixel by pixel correlation of the map of $f^{0.12}_{\rm low}\,\NHI^*$ with the CNM column density map obtained using \ROHSA\ (figure 7 upper left of \citealt{vujeva_2023}). Both axes are in units $10^{19}$\,cm$^{-2}$. The values from the FT method are clearly below the 1:1 line, by amounts bounded by the other lines discussed in the text.
  }
  \label{fig:LLIV1scatter}
\end{figure}

Visual comparison of these two maps reveals that the FT method retains information about the projected spatial structure, just as we found using simulated data in Section \ref{subsec:simu-result}. However, because $f^{0.12}_{\rm low}$ is a lower limit, the column density is lower.
This is quantified in Figure \ref{fig:LLIV1scatter}, a 2D histogram correlating the values in the two maps.

The CNM clusters in the \ROHSA\ $\sigma - \mu$ diagram are centered near 2 \kms.  If all CNM Gaussians had this dispersion, then their $f^{0.12}_{\rm low}$ would give the slope of dark blue line (second lowest), absent destructive interference effects. However, the clusters have some spread.  Some Gaussians are as broad as $\sigma = 3$ \kms, leading to slope 0.05 (lowest line), the fiducial value discussed above, and some are narrower, a few as small as 1 \kms. These lines nicely bracket the 2D histogram in Figure \ref{fig:LLIV1scatter}.

At low $N^*_{\mathrm{HI_{CNM,\, ROHSA}}}$ ($\lesssim$ 2$\times10^{19}$\,cm$^{-2}$), 
%\ROHSA\ CNM column density 
the plateau of low but finite $f^{0.12}_{\rm low}\,\NHI^*$ (corresponding to the darker regions in Figure~\ref{fig:LLIV1_CNM}) arises from map areas with very small values of $f^{0.12}_{\rm low}$ (typically below 0.05) from broad WNM emission and noise multiplied by significant $\NHI^*$ from WNM gas.

\section{Application to mapped 21\,cm data for DHIGLS DF}
\label{sec:DHIGLS}

The main science goal illustrated by the applications below is to search for regions where there has been a thermal phase transition with condensation of cold dense gas.  Searches in the future will need to be done over larger and larger data sets and in different spectral ranges and so require an efficient approach, such as we have developed in our FT-based method.  Interesting regions found can then be targeted for analysis with more compute-intensive methods like \ROHSA\ \citep{marchal_2019} to quantify the properties of the multiphase medium.

We have used the FT method on data from GHIGLS and DHIGLS for a rapid evaluation of other regions and velocity ranges to study.  The reader can explore this with the notebook supplied (footnote \ref{foot:notebook}). Two applications are described below, here and in Section \ref{sec:hi4pi}.

\subsection{Data} 
\label{subsec:data}
%DF cube and NCPL cube from DHIGLS and GHIGLS
The 57.4 square degree DF dataset used in the application here, located at $(\alpha,\, \delta) = (10^{{\mathrm h}}$30$^{{\mathrm m}},\, 73\degree 48')$, was produced in the DHIGLS\footnote{DRAO \HI\ Intermediate
Galactic Latitude Survey: \url{https://www.cita.utoronto.ca/DHIGLS/}} \HI\ survey \citep{blagrave_dhigls:_2017} with the Synthesis Telescope (ST) at the Dominion Radio Astrophysical Observatory.
The 256-channel spectrometer, spacing $\Delta v=0.824$\,\kms\  sampling data with velocity resolution 1.32\,\kms, was centered at $v_c=-60$\,\kms\ relative to the Local Standard of Rest (LSR). 
The spatial resolution of the ST interferometric data was about 0\farcm9.
DF, which encompasses the Spider region (named for its prominent ``legs" emanating from a central ``body": \citealt{mandm2023} and references therein), is embedded in the North Celestial Pole Loop (NCPL) mosaic of the GHIGLS \HI\ survey \citep{martin_ghigls:_2015} with the Green Bank Telescope (GBT), with spatial resolution about $9\farcm 4$. 
The DHIGLS DF product has the full range of spatial frequencies, obtained by
a rigorous combination of the ST interferometric and GBT single dish data
\citep[see section 5 in][]{blagrave_dhigls:_2017}. The pixel size is 18\arcsec.
A map of the total column density in the DF field in the velocity range $-15 < v < 15$\,\kms\ is shown in the Figure~\ref{fig:NHI_TOT_fft_LVC_DF}.

\subsection{Apodization} 
\label{subsec:apod}
Real measurements of the 21\,cm line frequently include emission from intermediate velocity clouds (IVCs) that overlap the local velocity component (LVC). For example in the NCPL, there is also significant emission bridging velocities in between, which does not show the loop and has some characteristics of IVC, but was distinguished as a Bridge Velocity Component (BVC) by \citet{taank_2022} in their analysis of the multiphase structure of the loop.
Building on this, we used $T_b$ in the velocity range $-15 < v < 15$\,\kms, effectively setting $T_b$ to zero outside of this range.
Introducing this sharp edge (step function) in the signal produces a ``ringing" in the Fourier transformed spectrum (Gibbs phenomenon). However, we suppress this by applying a cosine apodization on four channels of the signal at both ends of the spectral range.

\subsection{Denoising} 
\label{subsec:denoising}

\begin{figure}
  \centering
  \includegraphics[width=\linewidth]{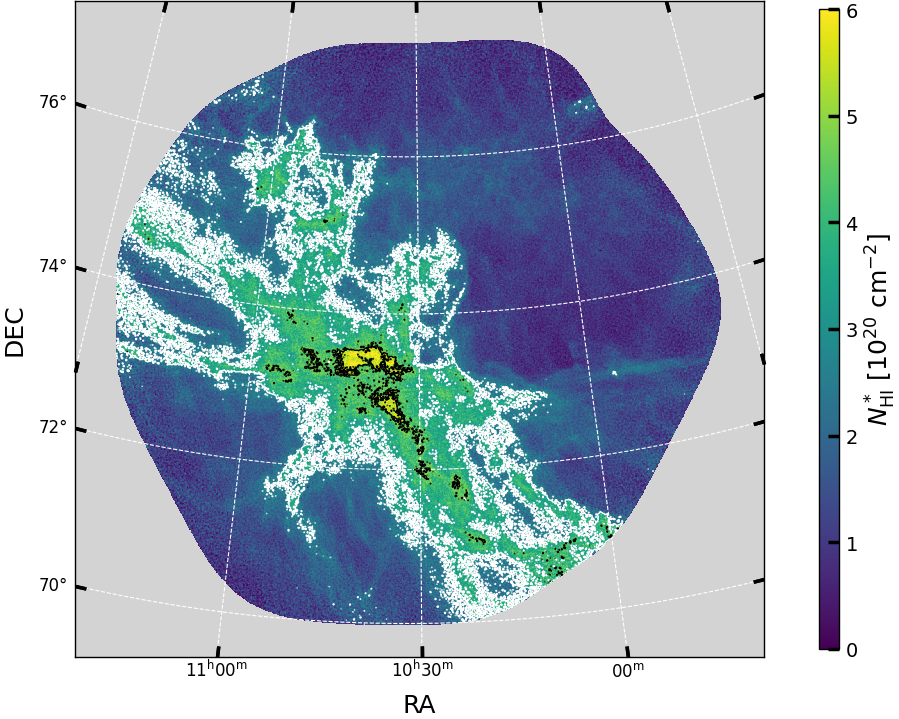}
  \caption{Total column density $\NHI^*$ of the DF dataset from DHIGLS in the velocity range $-15 < v < 15$\,\kms, computed in the optically thin limit.
  The solid white and black lines show the $\NHI^*=3\times10^{20}$\,cm$^{-2}$ and $\NHI^*=5\times10^{20}$\,cm$^{-2}$ contours.
  }
  \label{fig:NHI_TOT_fft_LVC_DF}
\end{figure}

\begin{figure}
  \centering
  \includegraphics[width=\linewidth]{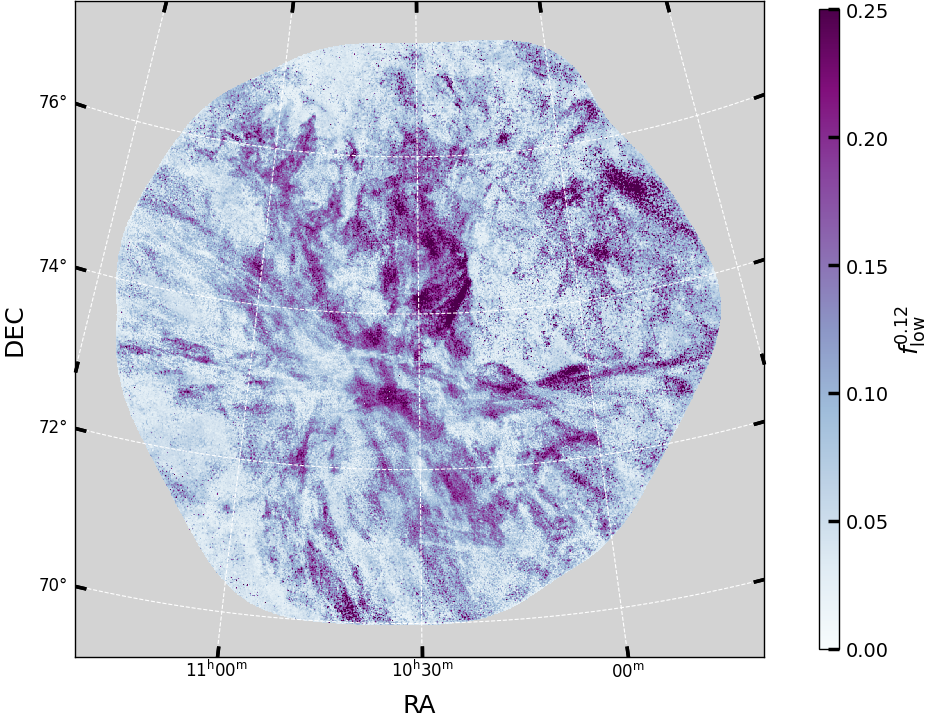}
  \includegraphics[width=\linewidth]{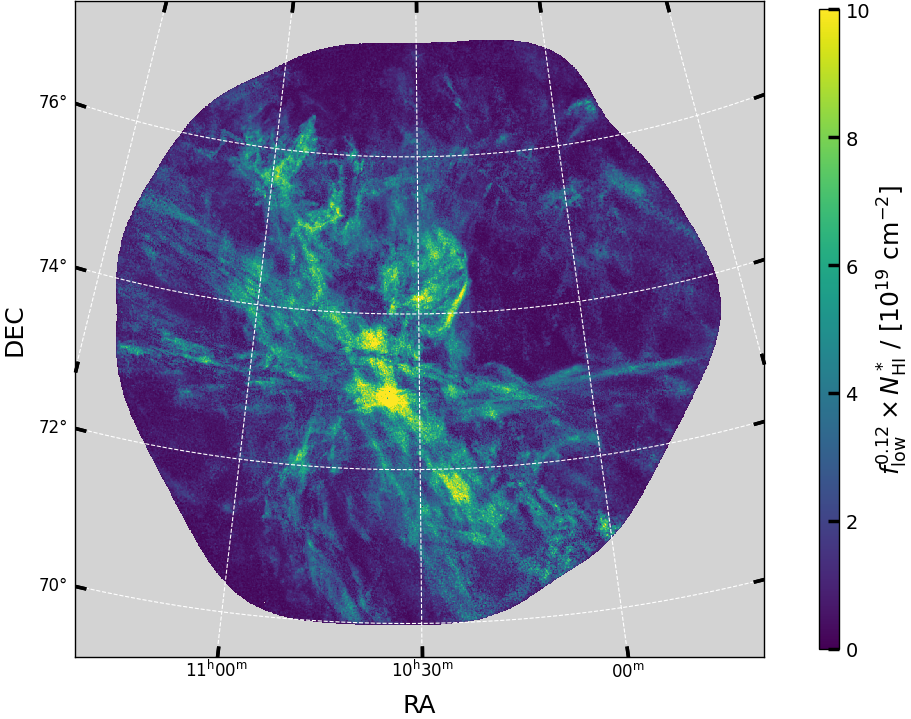}
  \caption{Lower limit on the cold gas mass fraction $f^{0.12}_{\rm low}$ (top) and corresponding lower limit on the cold gas column density $f^{0.12}_{\rm low}\,\NHI^*$ (bottom) of the DF dataset from DHIGLS in the velocity range $-15 < v < 15$\,\kms.
  }
  \label{fig:fcnm_fft_LVC_DF}
\end{figure}

Evaluation of the cold gas mass fraction using the method described in 
Section~\ref{subsec:lower-limit} is sensitive to noise. 
This can be appreciated in Figure~\ref{fig:diffraction_21Sponge} where noise dominates the signal at high $k_{v,\rm lim}$ values in the amplitude spectrum of the FT of the interpolated $T_b$ spectrum (black) of J2232 from 21-SPONGE.
To overcome this limitation, we used the \ROHSA\ algorithm \citep{marchal_2019} to ``denoise'' the whole hyper-spectral cube of the DF dataset. 

\ROHSA\ is a regularized optimization algorithm that decomposes position-position-velocity (PPV) data cubes into a sum of Gaussians \citep{marchal_2019}. {\tt ROHSA} takes into account the spatial coherence of the emission and its multi-phase nature to perform a separation of different thermal phases.
Earlier applications of {\tt ROHSA } were dedicated to mapping out thermal phases in 21\,cm data \citep[e.g., ][]{marchal_2021b,marchal_2021a,taank_2022,vujeva_2023}. 

Here we simply made use of the spatial regularization to obtain a spatially coherent model of the data that concurrently reduces noise.
The decomposition used in this work was obtained using $N=6$ Gaussians and the hyper-parameters $\lambda_{\ab}=\lambda_{\mub}=\lambda_{\sigmab}=\lambda_{\sigmab'}=10$, which control the strength of the regularization that penalizes small-scale spatial fluctuations of each Gaussian parameter (amplitude, central velocity, dispersion), measured by the energy at high spatial frequencies ($\lambda_{\ab}=\lambda_{\mub}=\lambda_{\sigmab}$), and the regularization of the variance of each velocity dispersion field ($\lambda_{\sigmab'}$).
This decomposition reproduces the original data with a good $\chi^2$ and we refer to the PPV cube built from the model as ``denoised data.''\footnote{This simple decomposition is good enough for denoising DF, then applying the FT method. However, it has not undergone the rigorous analysis and testing needed to recommend it for thermal phase separation in this complex field.}

Maps of the gas column density, 
$f^{k_{v,\rm lim}}_{\rm low}\,\NHI^*$,  as a function of $k_{v,\rm lim}$ computed from the denoised data and the original data are shown in Figures \ref{fig:hsfft_DF} and \ref{fig:hsfft_DF_brut}, respectively, in Appendix~\ref{app:denoising}.
At low $k_{v,\rm lim}$, the maps are similar but the relative proportion of noise compared to the sky signal increases with $k_{v,\rm lim}$ to a point where filamentary structures are hardly visible in Figure \ref{fig:hsfft_DF_brut}. Decomposition with {\tt ROHSA} concurrently reduces noise at high $k_{v,\rm lim}$, retaining spatial information about the very cold structures in the Spider in Figure \ref{fig:hsfft_DF}.

%Comment on when to apply a denoising
Application of a denoising algorithm should be considered based on the maximum kinetic temperature one wishes to probe in the data following the methodology described in Section~\ref{subsec:lower-limit} pairing $\sigma_{\rm lim}$ and $t_{\rm lim}$ to motivate the choice of $k_{v,\rm lim}$. 
Because the FT is computationally inexpensive (see Section~\ref{sec:discussion}) compared to denoising, we recommend exploratory application of the FT method without denoising the data to judge the quality of $f^{k_{v,\rm lim}}_{\rm low}$ at the selected $k_{v,\rm lim}$.  In our application to Spider with $f^{k_{v,\rm lim}}_{\rm low} = 0.12$, denoising the data is not particularly beneficial for most of the map, except for areas such as the upper right where the column density is low or around the perimeter where the noise is larger.  In those areas,  image is less noisy and the contrast between green structure and blue is enhanced.

\subsection{Results}
\label{subsec:result-obs}
The top panel of Figure~\ref{fig:fcnm_fft_LVC_DF} shows $f^{0.12}_{\rm low}$ (corresponding to $k_{v,\rm lim}=0.12$\,(\kms)$^{-1}$) in DF using the apodized denoised data. The bottom panel of Figure~\ref{fig:fcnm_fft_LVC_DF} shows the corresponding lower limit on the column density of cold gas, calculated by multiplying the total column density (Figure \ref{fig:NHI_TOT_fft_LVC_DF}, computed in the optically thin limit) by the cold gas mass fraction (top panel), i.e., $f^{0.12}_{\rm low}\,\NHI^*$.
Both panels reveal structure attributable to cold gas in the Spider, arranged in a complex network of ``filaments.''

Because $f^{0.12}_{\rm low}$ is a lower limit, we cannot discuss the exact spatial distribution of the cold gas, even in projection let alone 3D. However, our tests in Section~\ref{subsec:simu-result} n realistic simulations of HI gas show that information about the projected spatial structure is retained in the FT-based map. Here, applied to actual data, it seems unlikely that the structure that appears filamentary in projection is created at random; rather it reveals an important structural property of the thermal condensation of the \HI gas.

In the top panel, the lower limit $f^{0.12}_{\rm low}$ varies considerably across the field, ranging from 0 (no CNM) up to 0.3 along the filamentary structures. In the bottom panel, the brightest part of the lower limit on cold gas column density is located near the core of the Spider identified in the total column density map, but the contrast is higher in this map and the structural details differ.

For completeness, we present lower-limit gas mass fraction maps and column density maps for the first nine $k_{v,\rm lim}$ values obtained from the FT of the denoised data, in Figures~\ref{fig:f_hsfft_DF} and \ref{fig:hsfft_DF}, respectively.\footnote{Note that at high $k_{v,\rm lim}$, denoising the data prior to the FT becomes critical, as is evident in Figure \ref{fig:hsfft_DF_brut}, which shows the same column density maps as Figure \ref{fig:hsfft_DF} but computed with the original data.}
The PPK cube reveals increasingly colder structures as $k_{v,\rm lim}$ increases
and it can be appreciated visually that at high $k_{v,\rm lim}$ there is more structure on small spatial scales. Note that we applied a Fourier transform only along the velocity axis of the PPV cube, with no explicit spatial filtering. The relationship between small spatial scale features and the intensity at high $k_{v,\rm lim}$ can be understood as a direct consequence of the thermal condensation of \HI gas.

The network of filaments seen in $f^{0.12}_{\rm low}\,\NHI^*$ is similar to the filamentary structures revealed by scanning through the PPV cube of the original data. These structures were first illustrated by \citet{blagrave_dhigls:_2017} in their RGB color image of the \HI\ emission using three distinct channels separated by only 2.47\,\kms\ (top panel of their figure 24).
The dramatic changes in the filamentary structure of \HI\ channel maps over a small velocity separation is captured at high $k_v$ in the Fourier transformed data, hence the similarities between $f^{0.12}_{\rm low}\,\NHI^*$ and the structure of individual channel maps.
The FT approach and the channel maps provide complementary perspectives on the nature of filaments \citep[also called fibers in the context of the diffuse sky at high latitude,][]{clark_2014} that are believed to be dominated by cold density structures \citep{clark_2019,peek_clark_2019,murray_2020}.

This qualitative result is consistent with the recent spatial wavelet scattering transform (ST) analysis performed by \cite{lei_2023}. Using a combination of GALFA-HI data \citep{peek_2011,peek_2018} and 21-SPONGE data, the authors found that ST-based metrics for small-scale linearity are predictive of the fraction of cold gas along the line of sight. It is noteworthy that the ST approach is based solely on analyzing the multi-scale information of \HI\ emission maps while our FT method is solely applied on the spectral axis of the data cube. This complementarity opens up an interesting avenue for combining both the spatial and spectral information that can be extracted from 21\,cm data.

A comprehensive statistical characterization of the correlation between $f^{0.12}_{\rm low}\,\NHI^*$ (or different $k_{v,\rm lim}$) and the structure of individual channel maps is beyond the scope of this paper. This could be achieved using a cross correlation analysis.
Similarly, a thorough analysis of the whole PPK cube of DF is beyond the scope of this paper but in the future will provide valuable information about the multiphase and multi-scale structure of \HI\ gas in the Spider that is part of the NCPL.

\section{Application to HI4PI data for the whole sky} 
\label{sec:hi4pi}
%\am{Add note with zonedo (or another data center) to share the healpix HI4PI maps.}

Using the HI4PI data (Section \ref{sec:validation-hi4pi}), we computed $f^{0.12}_{\rm low, HI4PI}$ in the velocity range $-90 < v < 90$\,\kms\ over the full sky. 
No denoising was applied to the dataset prior to the Fourier transform because the noise level of $T_b$ is fairly uniformly low, about $\sigma_{\rm rms}\sim43$\,mK.
Each spectrum was apodized with a cosine function at both ends of the spectral range.

\begin{figure*}
  \centering
  \includegraphics[width=\linewidth]{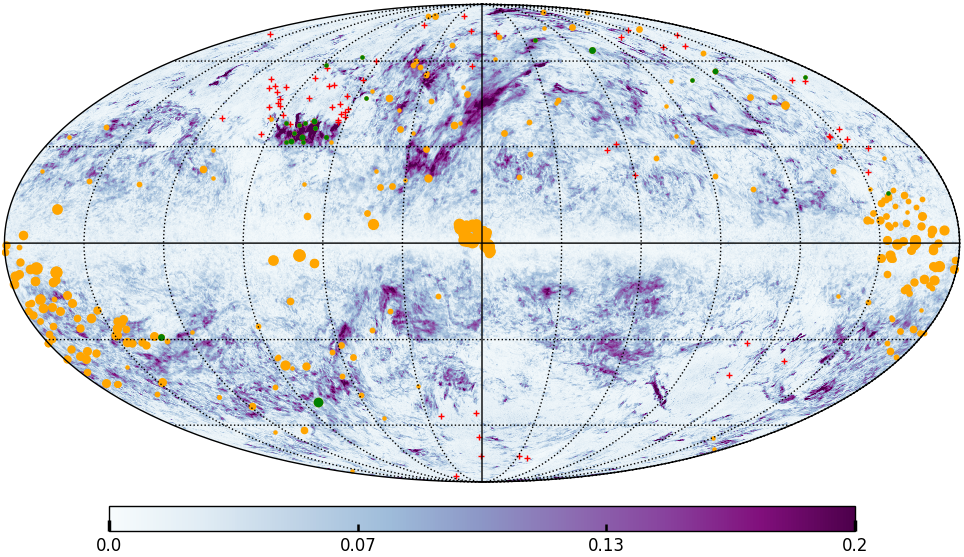}
  \includegraphics[width=\linewidth]{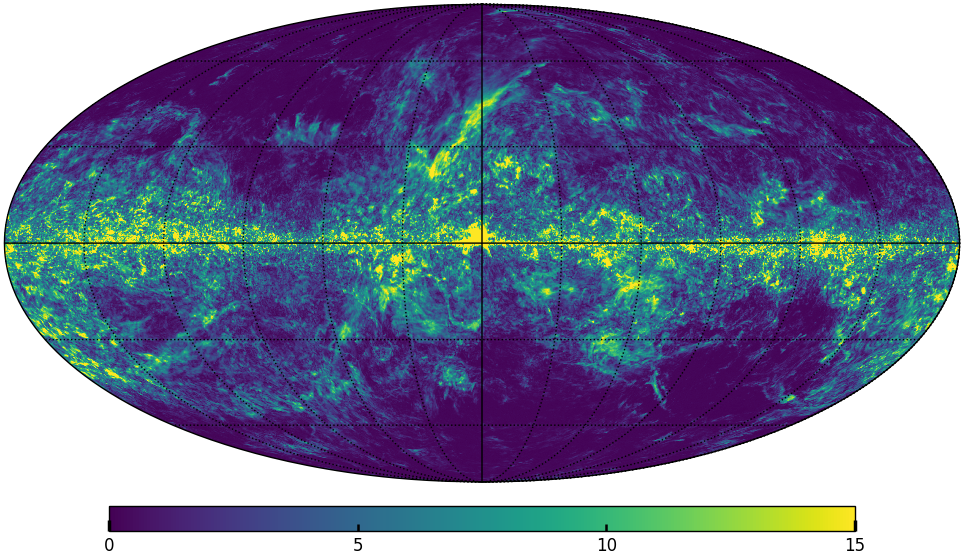}
  \caption{Mollweide projection centered on the Galactic center of the lower limit on the cold gas mass fraction $f^{0.12}_{\rm low, HI4PI}$ (top) and corresponding lower limit on the cold gas column density $f^{0.12}_{\rm low, HI4PI}\,\NHI^*$ in units of 10$^{19}$\,cm$^{-2}$ (bottom) of the entire HI4PI dataset in the velocity range $-90 < v < 90$\,\kms.
  Similar to the annotations in Figure~\ref{fig:NHI_TOT_LOWER_HI4PI_mollview}, the red crosses in the top panel show the positions of non-detections in the compilation of \HI\ absorption spectra from the BIGHICAT meta-catalog \citep{mcclur23} 
  and the orange dots, whose size encodes the number of absorption components (ranging from 1 to 16) in their $\tau$ spectrum, show the detections in BIGHICAT.  %
 The green dots show where detections in BIGHICAT have $f_{\rm BIGHICAT} - f^{0.12}_{\rm low, HI4PI} < 0$.
  }
\label{fig:f_CNM_LOWER_HI4PI_mollview}
\end{figure*}

\begin{figure}
  \centering
  \includegraphics[width=\linewidth]{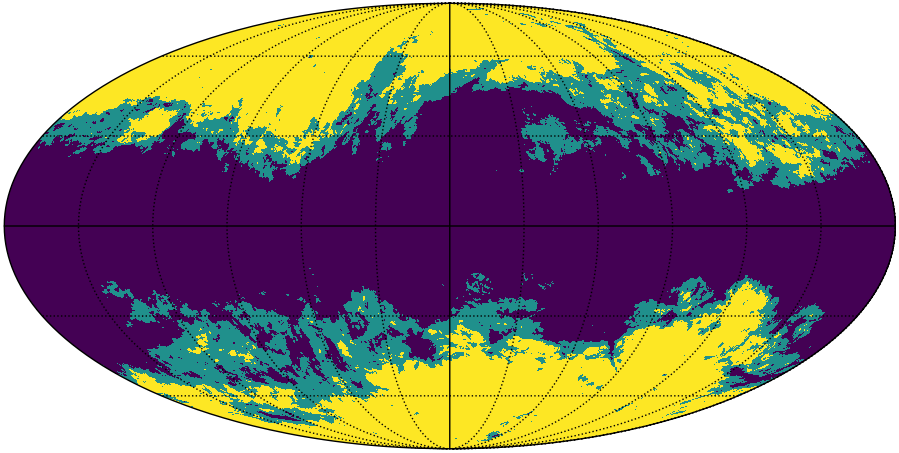}
  \caption{Mollweide projection centered on the Galactic center showing regions with $\NHI^* < 3\times10^{20}$\,cm$^{-2}$ (yellow), $ 3\times10^{20} < \NHI^* < 5\times10^{20}$\,cm$^{-2}$ (green), and $\NHI^* > 5\times10^{20}$\,cm$^{-2}$ (blue) from the HI4PI survey over the velocity range $-90 < v < 90$\,\kms.}
\label{fig:mask_3_5_HI4PI_mollview}
\end{figure}

Figure~\ref{fig:f_CNM_LOWER_HI4PI_mollview} shows Mollweide projections centered on the Galactic center of $f^{0.12}_{\rm low, HI4PI}$ (top) and $f^{0.12}_{\rm low, HI4PI}\,\NHI^*$ in units of 10$^{19}$\,cm$^{-2}$ (bottom). 
Annotations showing results from BIGHICAT are similar to those in Figure~\ref{fig:NHI_TOT_LOWER_HI4PI_mollview}.\footnote{For completeness, in Figure~\ref{fig:FMIN_CNM_LOWER_HI4PI_mollview} we present the same two quantities but for nine $k_{v,\rm lim}$ values sampled in the range $0.01<k_v\,(\kms)^{-1} < 0.27$.}
Comparison of the top map of $f^{0.12}_{\rm low, HI4PI}$  with maps made with increasing values of $k_{v,\rm lim}$ above 0.12 (\kms)$^{-1}$ (Figure \ref{fig:FMIN_CNM_LOWER_HI4PI_mollview} top in Appendix \ref{app:hi4pi}) reveals the same morphology but with amplitude decreasing with $k_{v,\rm lim}$, showing that the features detecting CNM gas are real, not noise dominated.

At high latitudes, these full sky maps show a wide variety of known regions, e.g., the North Polar Spur \citep[][and reference within]{tummer_1958,hanbury-brown_1960,das_2020,west_2021}, the compressed magnetized shells associated to the Corona Australis molecular cloud \citep{bracco_2020}, or the North Celestial Pole Loop \citep{meyer91,taank_2022,mandm2023}. 
A comprehensive description is beyond the scope of this paper and will be explored in future works.

Toward the Galactic plane, $f^{0.12}_{\rm low, HI4PI}$ is low, but due to the high total column density (computed in the optically thin limit), $f^{0.12}_{\rm low, HI4PI}\,\NHI^*$ shows a variety of structures with column density comparable to that of the high latitude sky.
At these latitudes, a sum of Gaussians as a model to describe the 21\,cm line is not a good description of the data, due to the extreme effect of opacity and self-absorption. The Riegel-Crutcher cloud seen toward the Galactic center \citep{heeschen_1955,riegel_1972,mcclure_2006} is a noteworthy example. \HI\ self-absorption (HISA) seen in its emission spectrum \citep[see, e.g., figure 2 in][]{mcclure_2006} is most likely to impact the amplitude structure of $\hat T_b(k_v)$. 
Narrow emission lines in the optically thin limit and HISA will both be highlighted at high $k_v$ in the Fourier transform of the emission data but whether Equation~\ref{eq:mass_fraction} still provides a lower limit on the amount of cold gas remains unclear and should be investigated in future work. We therefore recommend that $f^{0.12}_{\rm low, HI4PI}$ at very low Galactic latitude be taken with caution, especially in places where HISA are known to impact strongly the emission lines.

Figure~\ref{fig:mask_3_5_HI4PI_mollview} shows masks of regions with $\NHI^* < 3\times10^{20}$\,cm$^{-2}$ (yellow), $ 3\times10^{20} < \NHI^* < 5\times10^{20}$\,cm$^{-2}$ (green), and $\NHI^* > 5\times10^{20}$\,cm$^{-2}$ (blue). In the green and yellow areas (mainly the high latitude sky), where $f_{\rm BIGHICAT}$ and $f^{0.12}_{\rm low, HI4PI}$ are fairly similar, our lower limit estimate tends toward the absolute value of cold gas mass fraction probed by absorption line surveys (see discussion of Figure~\ref{fig:fcnm_fft_vs_BIGHICAT_high_b_er_NHI_thick} in Section \ref{sec:validation-bighicat}). In the blue areas, our lower limit should be treated as its definition, saying the cold gas mass fraction is at least $f^{0.12}_{\rm low, HI4PI}$ but its absolute value remains unknown from this method.
We recommend using these column density cuts to determine regions of the sky where $f^{0.12}_{\rm low, HI4PI}$ tends towards the absolute value of the cold gas mass fraction and where $f^{0.12}_{\rm low, HI4PI}$ should be strictly used as a lower limit.

\section{Assessing \lowercase{$f^{0.12}_{\rm low}$} in the high latitude sky using a CNN}
\label{sec:validationhighlat}

\begin{figure*}
  \centering
  \includegraphics[width=0.8\linewidth]{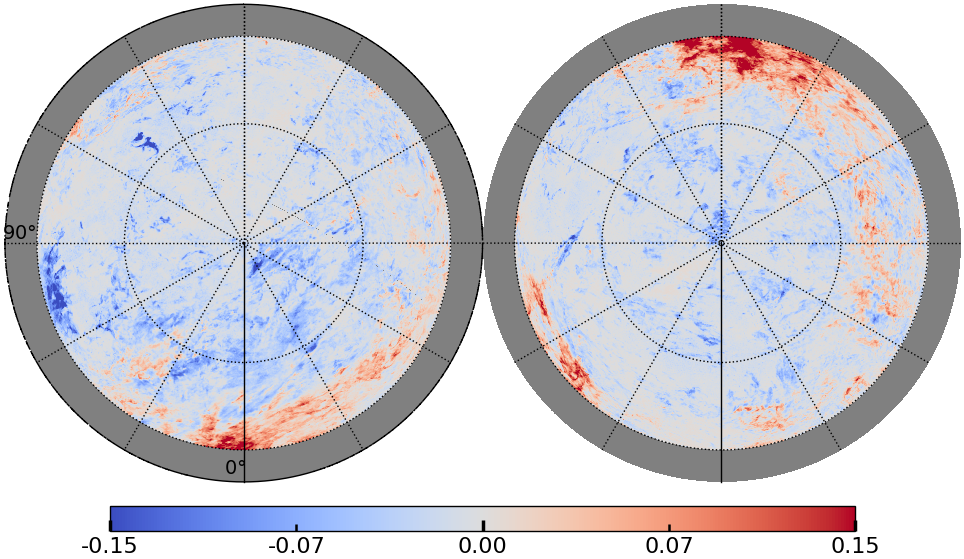}
  \caption{Orthographic projection of the northern (left) and southern (right) Galactic hemisphere of $f_{\rm CNN}^{\rm HI4PI} - f^{0.12}_{\rm low, HI4PI}$. For orientation, $\ell=0\degree$ and $\ell=90\degree$ are annotated on the left panel.\am{fixme}
  }
  \label{fig:Diff_FFT_CNN_HI4PI_mollview}
\end{figure*}

\begin{figure}
  \centering
  \includegraphics[width=\linewidth]{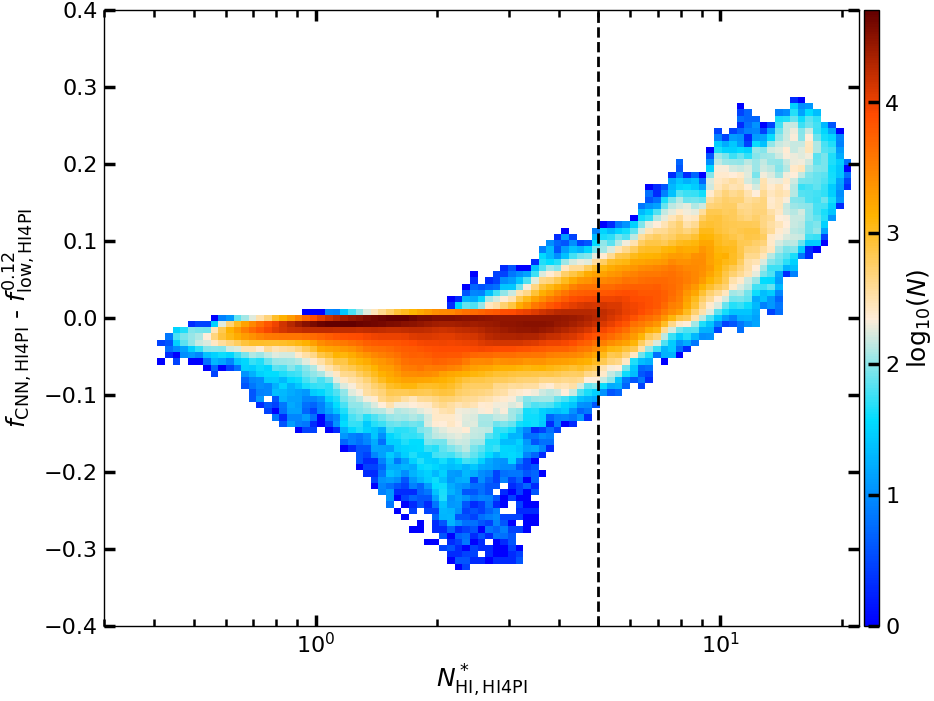}
  \caption{2D distribution function of $f_{\rm CNN}^{\rm HI4PI} - f^{0.12}_{\rm low, HI4PI}$ vs.\ $\log \NHI^*$ computed in the optically thin limit with HI4PI data.
  The dashed vertical line denotes the typical column density where the effects of opacity become significant, as in Figure\ref{fig:fcnm_fft_vs_21sponge_high_b_er_NHI_thick_color_b}.
  }
  \label{fig:heatmap_CNM_FFT}
\end{figure}

\citet{hensley_2022} applied the CNN model developed by \cite{murray_2020} to HI4PI data in the same velocity range, $-90 < v < 90$\,\kms, but only in the high latitude sky (i.e., $|b|>30\degree$),  obtaining a map of $f_{\rm CNN}^{\rm HI4PI}$ in both Galactic hemispherical caps.

The two estimates of the cold mass gas fraction are compared in 
Figure~\ref{fig:Diff_FFT_CNN_HI4PI_mollview}, an orthographic projection of the difference $f_{\rm CNN}^{\rm HI4PI} - f^{0.12}_{\rm low, HI4PI}$ in northern (left) and southern (right) Galactic hemispheres.
Figure \ref{fig:heatmap_CNM_FFT} 
shows the 2D distribution function of the difference vs.\ $\log \NHI^*$.

The two estimates are in fairly good agreement, with the mean and standard deviation of the difference $f_{\rm CNN}^{\rm HI4PI} - f^{0.12}_{\rm low, HI4PI}$ being $-0.0065$ and 0.033, respectively. 
However, there are coherent regions of positive and negative excursions around the mean, predominately in the range $30<|b|\degree<60$.
Positive values are predominantly in regions of higher column density, where the CNN, through its training, performs better at correcting for opacity effects \citep[see figure 8 in][]{murray_2020}.

Negative values are found in more localized clouds, prominent at, e.g., ($\ell, b$) = (75\degree, 35\degree) (which corresponds to the region covered by the MACH survey and is discussed in Section~\ref{sec:validation-bighicat}) and ($\ell, b$) = (135\degree, 55\degree).
The emission of the first cloud is in the low velocity range (LVC). The second cloud is a bright ($T_b\approx26$\,K) IVC at about $-50$\kms.\footnote{A $4^\circ x\, 4^\circ$ subset of this area was mapped in ICRS coordinates in the GHIGLS survey with the GBT at 9\farcm55 resolution. There it is called the UM1 field, centered at ($\ell, b$) = (134\fdg95, 54\fdg13) coincidentally near the peak of the IVC-dominated column density.}
The CNN and the FT methods have both been applied to the same emission spectra at 16\farcm2 resolution and so unlike for the comparison between our lower limit and the BIGHICAT estimate based on absorption, this discrepancy cannot be explained by the beam of the observation.

\begin{figure*}
  \centering
  \includegraphics[width=0.32\linewidth]{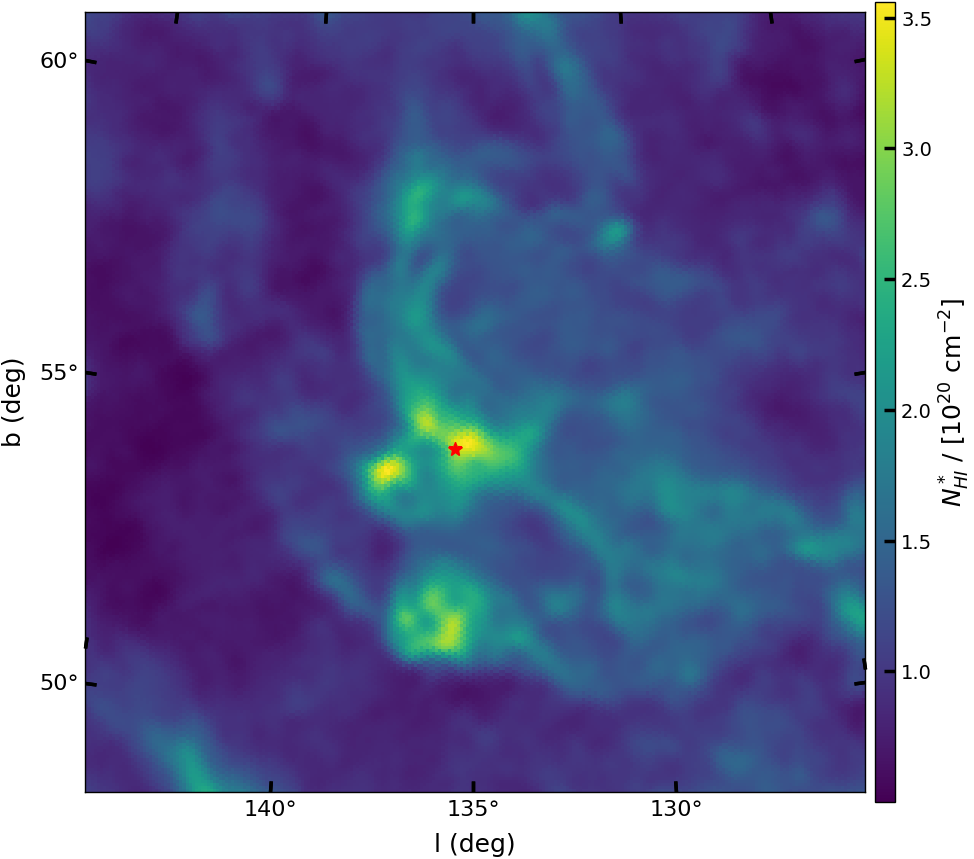}
  \includegraphics[width=0.32\linewidth]{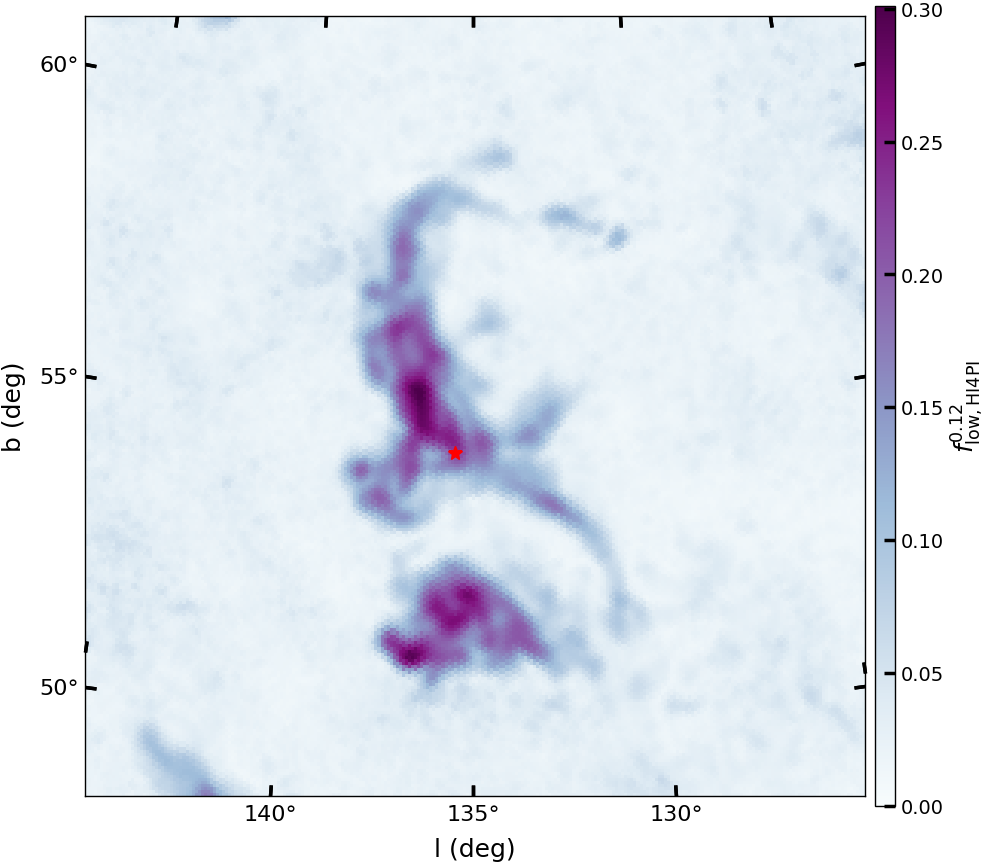}
  \includegraphics[width=0.32\linewidth]{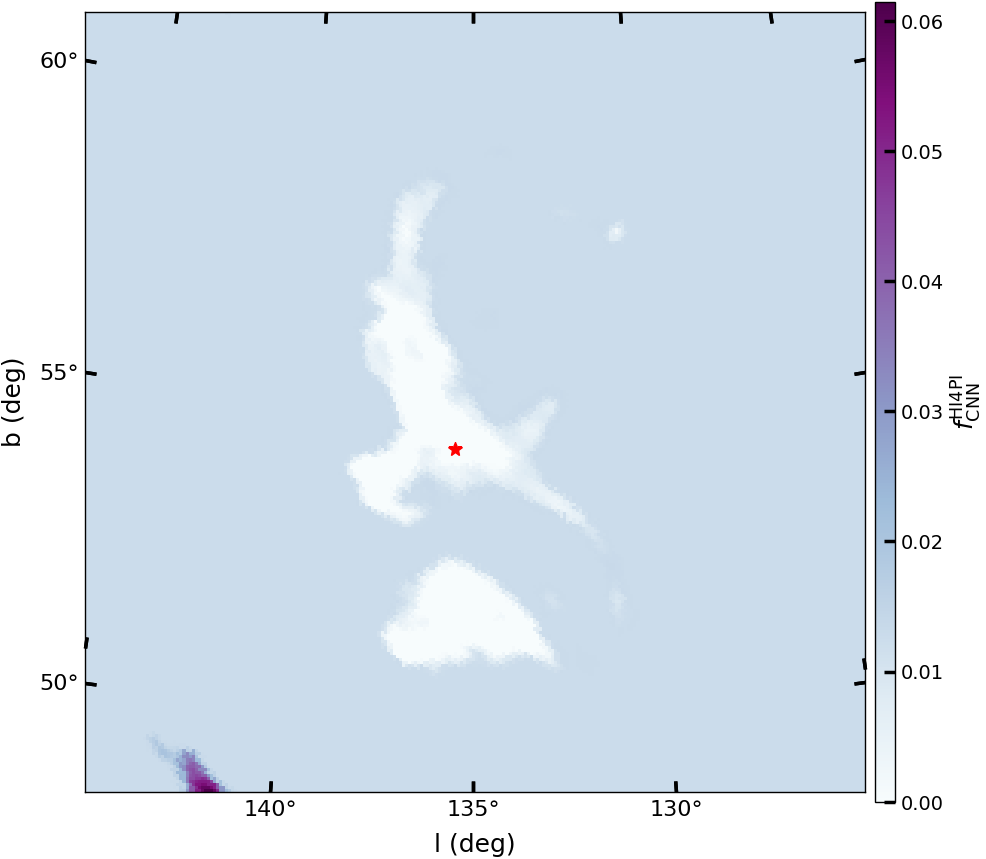}
  \caption{Left: Tangential projection in Galactic coordinates of $N^*_{HI}$ in the velocity range $-90 < v < 90$\,\kms\ from HI4PI, centered at ($\ell, b$) = (135\degree, 55\degree).
  Middle: $f^{0.12}_{\rm low, HI4PI}$. Right: $f_{\rm CNN}^{\rm HI4PI}$.
  The red star marks ($\ell, b$) = (136\fdg3, 54\fdg7), the line of sight of the spectrum whose Gaussian decomposition is shown in Figure~\ref{fig:Tb_decomposed}.
  }
  \label{fig:CNN}
\end{figure*}
\begin{figure}
  \centering
  \includegraphics[width=\linewidth]{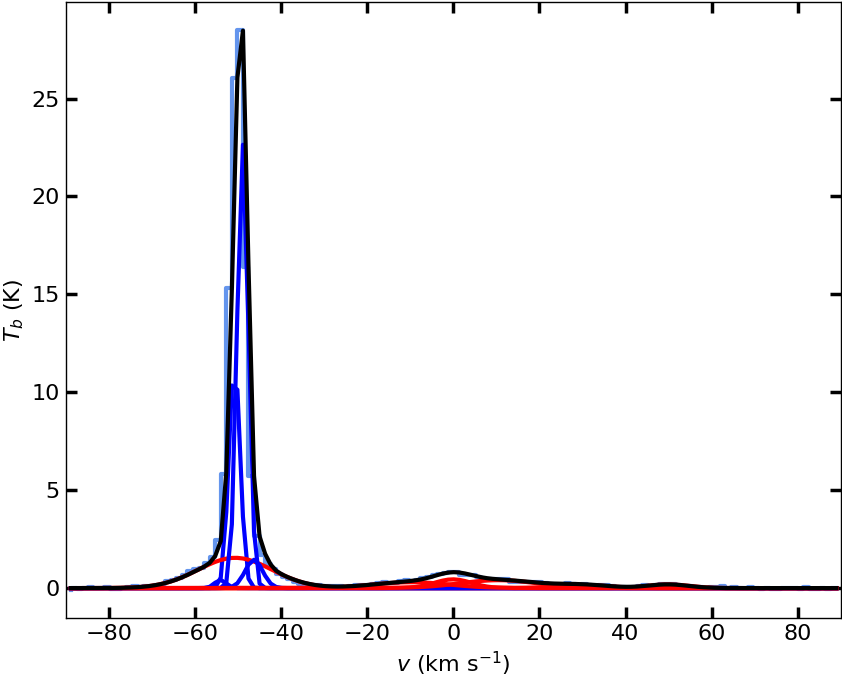}
  \caption{Brightness temperature spectrum extracted from HI4PI at ($\ell, b$) = (136\fdg3, 54\fdg7), located by the red star in Figure~\ref{fig:CNN}. 
  The data (light blue histogram) is well matched by the total model (black curve).
  Individual Gaussians with $\sigma < 3$\kms\ ($\sigma > 3$\kms) are shown in blue (red). The gas mass fraction in cold Gaussians evaluates to 0.63 and our lower limit estimate $f^{0.12}_{\rm low, HI4PI} = 0.28$.
  }
  \label{fig:Tb_decomposed}
\end{figure}

We have investigated the IVC cloud, where the difference between the two methods is the most negative.
Its column density map $N^*_{HI}$ in the velocity range $-90 < v < 90$\,\kms\ from HI4PI is shown in the left panel in Figure~\ref{fig:CNN}.
Figure~\ref{fig:CNN} (middle) shows a map of $f^{0.12}_{\rm low, HI4PI}$.
It has a structure similar to that of the total column density within the IVC cloud, a shape clearly recognized in the Figure \ref{fig:Diff_FFT_CNN_HI4PI_mollview}.
We have extracted a spectrum at ($\ell, b$) = (136.3\degree, 54.7\degree) in a bright sub-structure of the cloud (annotated by the red star in Figure~\ref{fig:CNN}) where $f^{0.12}_{\rm low, HI4PI}$ is quite high (0.28). Figure~\ref{fig:Tb_decomposed} shows this $T_b$ spectrum in light blue. 
We have decomposed the spectral data with a Gaussian model, where the number of Gaussian ($N=10$) was selected based on a reduced chi-square criterion.
The total model is shown in black.  Individual Gaussians with $\sigma < 3$\kms\ ($\sigma > 3$\kms) are shown in blue (red). 
The gas mass fraction in the cold Gaussians (blue) evaluates to 0.67, higher than our lower limit $f^{0.12}_{\rm low, HI4PI}$.

On the other hand, the map of the estimate $f_{\rm CNN}^{\rm HI4PI}$ (Figure~\ref{fig:CNN}, right) is overall very close to 0; while it also shows a structure of the same shape, its mean value is curiously negative, about -0.04, lower than that of the background of about 0.01. This failure of the CNN probably arises because the IVC-dominated spectra are far different than the training set used for the CNN. 
%This is an ongoing experiment done by Van Hiep here at RSAA.
In support of this, if we translate the IVC peak in the spectrum in Figure \ref{fig:Tb_decomposed} to be centered on 0 \kms, then the CNN method does detect CNM gas, with $f_{\rm CNN}^{\rm HI4PI}= 0.22$.
%\am{Hiep will produce this map (translated cube with Claire's CNN that he will retrain based on her publicly available notebook).}
% \pgm{Let's leave this a qualitative statement. Did you do this yourself enough to convince you, in which case no need to mention the new work.  If there is a public CNN notebook, then you can just shift one spectrum and see that there is CNN.  No need to do clever things over the whole sky.  If so, how much would Figure 27 and 28 be affected, etc.  We would not need anything beyond line 1127?? Sounds like a can of worms that does not matter to us.}

%Investigating why the CNN method failed to find any CNM in this cloud is beyond the scope of this paper but should be addressed in future work.

%
\section{Discussion}
\label{sec:discussion}
%Pro and cons of our approach
The FT method presented in this work has disadvantages as well as advantages.
By focusing on only the amplitude information of the PPK cube, the phase information is inevitably lost and only the average information about the amount of cold gas remains available to some extent. The impact of the velocity field on the amplitude of the FT (i.e., the interference patterns) constrains us to evaluating a lower limit on the mass fraction of cold gas. 

However, this partial loss of information of the velocity field, as well as the absence of modeling in comparison to Gaussian decomposition algorithms, makes our approach highly efficient in terms of computing time and so a unique tool to explore where the phase transition has taken place.
Note that the computing time is not reduced in the case where {\tt ROHSA} is used to denoise the data prior to the FT but, as discussed in Section \ref{subsec:denoising} and Appendix~\ref{app:denoising}, denoising only becomes critical at high $k_{v,\rm lim}$ and the cold gas mass fraction map inferred at $k_{v,\rm lim}=0.12$\,(\kms)$^{-1}$ using the original data is still practicable for further analysis.

As a reference, the PPK cube of the DF field with dimensions 60, 1680, and 1764 along the spectral and two spatial axes, respectively, was obtained in about a minute on a single CPU using the real fast Fourier transform algorithm implemented in {\tt NumPy}. 
The lower resolution full sky HI4PI data were Fourier transformed in about ten minutes on a single CPU.
The computation time scales linearly with the number of pixels because each line of sight is treated independently while applying the FT. This offers the possibility of straightforward parallelization on multiple CPUs. Furthermore, almost no additional memory is required as the operation can be performed independently for each spectrum.
Both the low computing time and memory make this method easily applicable to massive data sets, pointing the way to new approaches suitable for the new generation of radio interferometers like the Australian Square Kilometer Array pathfinder \citep[ASKAP, ][]{dickey_2013,pingel_2022}, the Square Kilometer Array \citep[SKA, ][]{mcclure_2015,de-blok_2015}, as well as the Next Generation Very Large Array \citep[ngVLA, ][]{pisano_2018}.

%Perspectives 
To overcome the loss of information of the velocity field and fully probe the fraction of cold gas as a function of velocity, it is necessary to go beyond the statistics encoded in the amplitude of the FT. Future research ought to examine the Short-Time Fourier Transform (STFT) and/or wavelet transform, which extract information from the $k_v$ domain while preserving information in the time domain (i.e., here the velocity domain), to facilitate progress on understanding the multiphase (multi-scale in terms of velocity) nature of the medium traced by 21\,cm data.

\section{Summary} 
\label{sec:summary}

We develop a new method for spatially mapping a lower limit on the mass fraction of the cold neutral medium by analyzing the amplitude structure of $\hat T_b(k_v)$, the Fourier transform of $T_b(v)$, the spectrum of the brightness temperature of \HI\ 21\,cm line emission with respect to the radial velocity $v$.
This development is supported by multiple experiments including a basic understanding of the FT spectrum using toy models, evaluation on a realistic simulation of the neutral ISM, validation using absorption line surveys, and application to real data.
The main conclusions are as follows.

\begin{itemize}
    \itemsep-0.2em
    \item Using toy models, we illustrate the origin of interference patterns seen in $\hat T_b(k_v)$. Building on this, a lower limit on the cold gas mass fraction is obtained from the amplitude of $\hat T_b$ at high $k_v$.

    \item Tested on a numerical simulation of thermally bi-stable turbulence, the lower limit from this method has a strong linear correlation with the ``true" cold gas mass fraction from the simulation for relatively low cold gas mass fraction. At higher mass fraction, our lower limit is lower than the ``true" value likely due to a combination of interference and opacity effects.

    \item Comparison with absorption surveys shows that our lower limit is close to the absolute value obtained from a combination of emission and absorption data for a total column density along the line of sight $\NHI\lesssim 3-5\times10^{20}$\,cm$^{-2}$, with a departure from linear correlation at higher column densities also due to a combination of interference and opacity effects.

    \item Application to the DRAO Deep Field (DF) from DHIGLS reveals a complex network of filaments in the Spider, an important structural property of the thermal condensation of the \HI gas. Our estimator $f^{0.12}_{\rm low}$ varies considerably across the field, ranging from 0 (no CNM) up to 0.3 along the filamentary structures. The brightest parts in the map of the lower limit on the cold gas column density are located near the core of the Spider identified in the total column density map.

    \item Application to the HI4PI survey in the velocity range $-90 < v < 90$\,\kms\ provides a full sky map of a lower limit on the mass fraction of the cold neutral medium that is fairly consistent with results from the CNN model produced by \cite{murray_2020}.
\end{itemize}

Although highly efficient in terms of computing time, the main limitation of our FT approach is the loss of information encoded in the velocity field. Future research ought to examine the Short-Time Fourier Transform and/or wavelet transform, which extract information from the $k_v$ domain while preserving information in the velocity domain.

\begin{acknowledgments}
We acknowledge support from the Natural Sciences and Engineering Research Council (NSERC) of Canada.
A.B. acknowledges support from the European Research Council through the Advanced Grant MIST (FP7/2017-2022, No.742719).
This research has made use of the NASA Astrophysics Data System. 
This work began under the program “Grand Cascade” organized and hosted by Institut Pascal at Université Paris-Saclay and the Interstellar Institute. We appreciate enlightening conversations with J.E.G. Peek, C. Murray, N. Pingel, and other members of the Interstellar Institute.
We appreciate enlightening conversations with Van Hiep Nguyen.
We thank the anonymous referee whose comments and suggestions have improved this manuscript.

\end{acknowledgments}

\software{Matplotlib \citep{hunter_2007}, {\tt NumPy} \citep{van_der_walt_2011}}

%%%-------------------------------------------------------
% --------- APPENDIX ------------------------------------------

% \pagebreak
\appendix
\counterwithin{table}{section}
\counterwithin{figure}{section}

\section{DHIGLS DF}
\label{app:dgihlsdf}

\subsection{Multiple $k_{v,\rm lim}$ limits}
\label{app:multikv}

\begin{figure*}
  \centering
  \includegraphics[width=\linewidth]{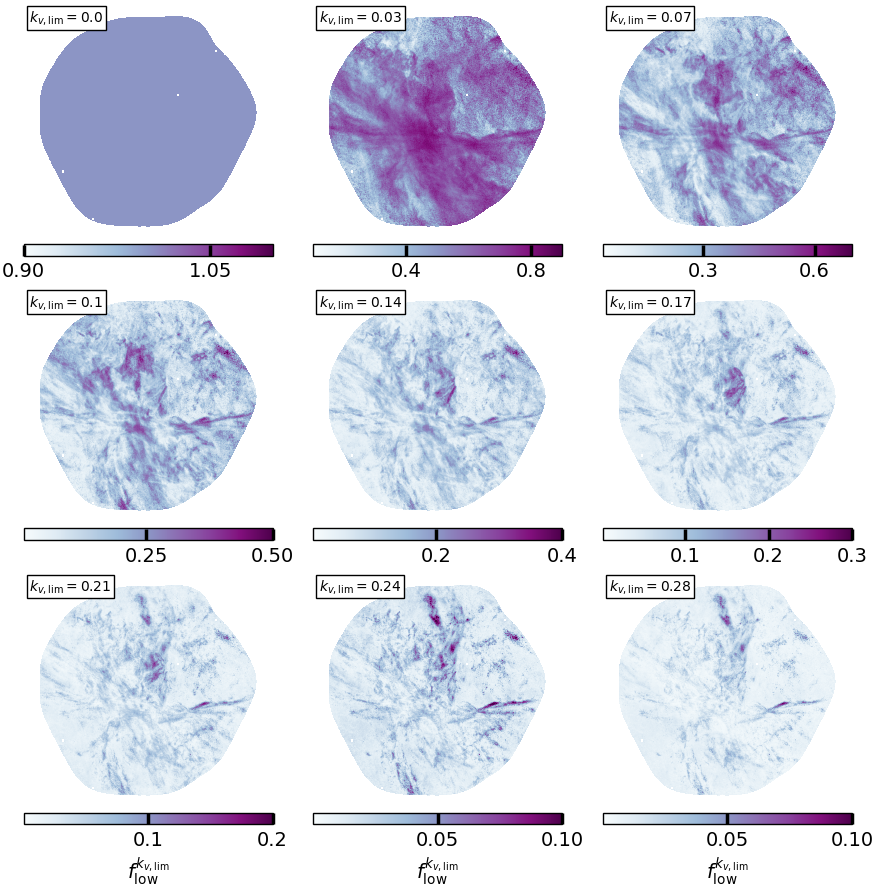}
  \caption{Gas mass fraction $f^{k_{v, \rm lim}}_{\rm low}$ as function of $k_{v,\rm lim}$ computed from the denoised data of the DF field from DHIGLS.}
  \label{fig:f_hsfft_DF}
\end{figure*}

\begin{figure*}
  \centering
  \includegraphics[width=\linewidth]{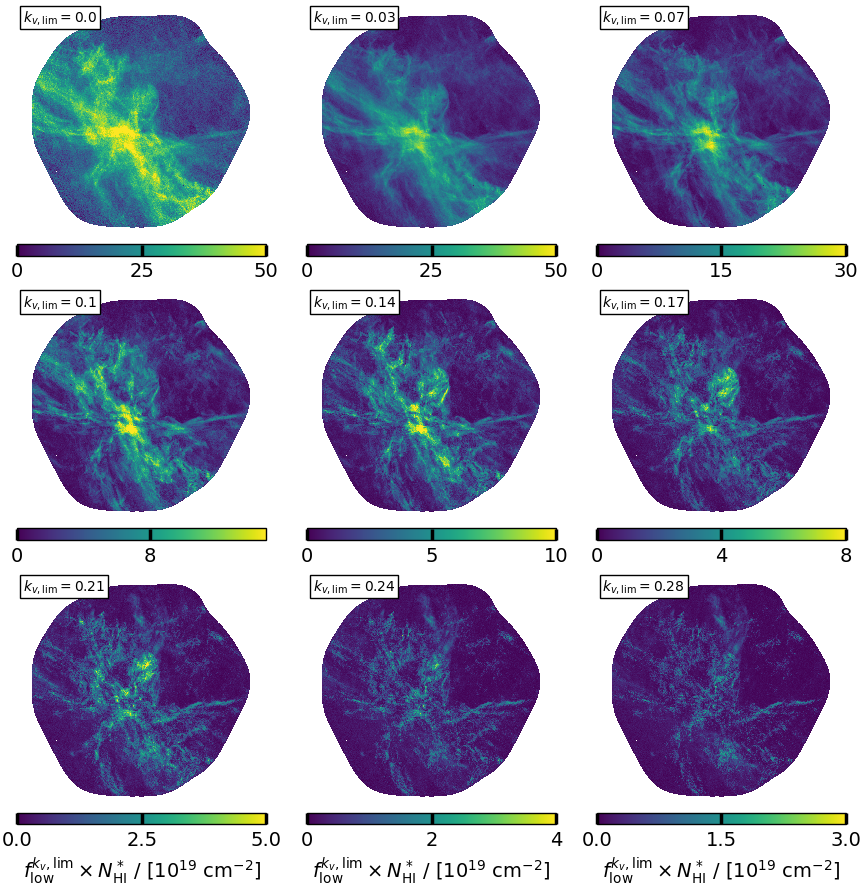}
  \caption{Column density $f^{k_{v,\rm lim}}_{\rm low}\,\NHI^*$ as function of $k_{v,\rm lim}$ computed from the denoised data of the DF field from DHIGLS.}
  \label{fig:hsfft_DF}
\end{figure*}

We present in Figures~\ref{fig:f_hsfft_DF} and \ref{fig:hsfft_DF}, respectively, the gas mass fraction maps and column density maps for the first nine $k_{v,\rm lim}$ values obtained from the FT of the denoised data of the DF field from DHIGLS.

\subsection{Denoising the data prior to the FT}
\label{app:denoising}

\begin{figure*}
  \centering
  \includegraphics[width=\linewidth]{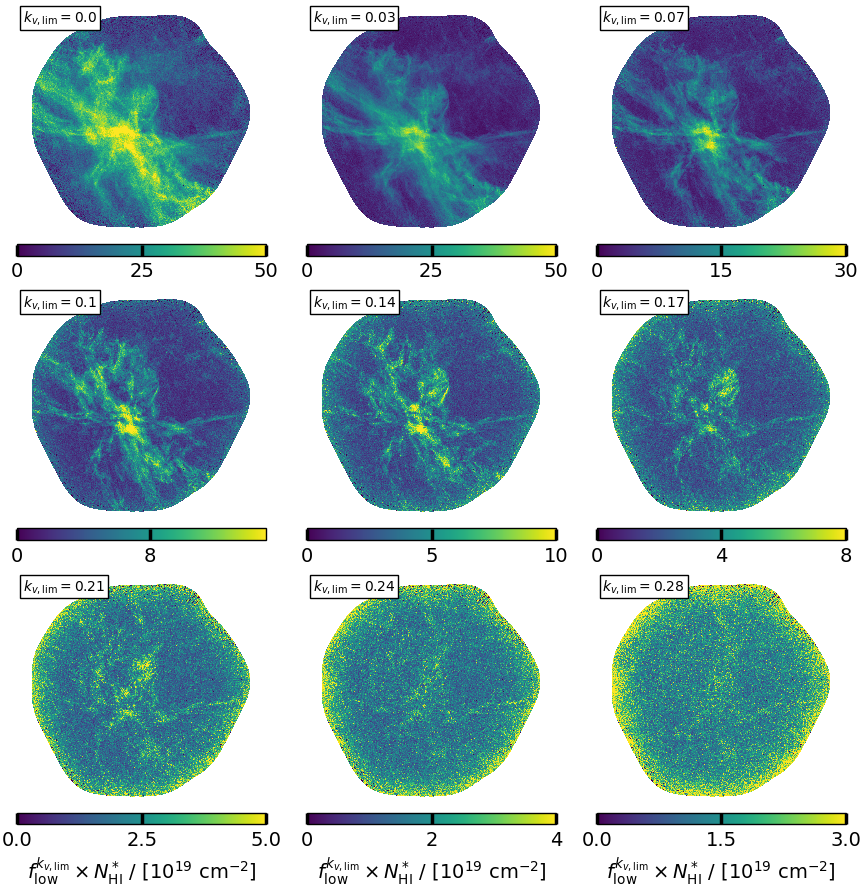}
  \caption{Column density $f^{k_{v,\rm lim}}_{\rm low}\,\NHI^*$ as function of $k_{v,\rm lim}$ computed from the original data (no denoising) of the DF field from DHIGLS.}
  \label{fig:hsfft_DF_brut}
\end{figure*}

To provide a baseline for evaluating the impact of the denoising of the DF data using {\tt ROHSA} on the PPK cube, we computed the FT of the original data. Figure~\ref{fig:hsfft_DF_brut} shows the column density $f^{k_{v,\rm lim}}_{\rm low}\,\NHI^*$  as a function of $k_{v,\rm lim}$. 
This can be compared panel by panel with Figure~\ref{fig:hsfft_DF} for the denoised data, presented with consistent color bars.

At low $k_{v,\rm lim}$, Figures~\ref{fig:hsfft_DF} and \ref{fig:hsfft_DF_brut} are very similar. However, proceeding toward higher $k_{v,\rm lim}$, the maps in Figure \ref{fig:hsfft_DF_brut} become increasingly noisier, up to a point where the filamentary structures are hard to distinguish from the ambient noise (see especially the map at the highest $k_{v,\rm lim}$).

\section{HI4PI}
\label{app:hi4pi}

\begin{figure*}
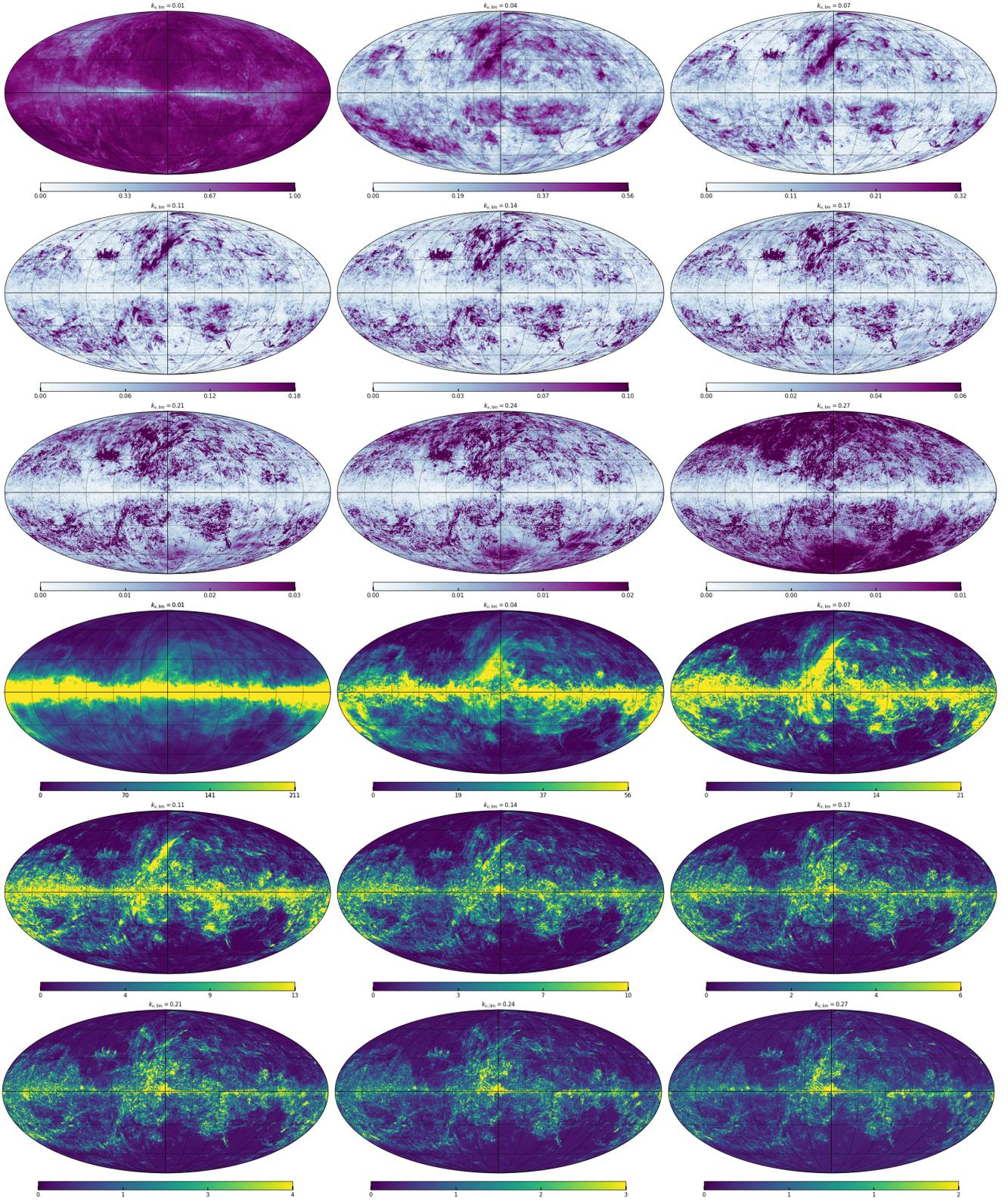

  \centering
  \foreach \i in {1,7,13,19,25,31,37,43,49} {%
  \includegraphics[width=0.32\linewidth]{figures/PPK_HI4PI/FMIN_CNM_LOWER_HI4PI_mollview_\i.png}
  }
  \foreach \i in {1,7,13,19,25,31,37,43,49} {%
  \includegraphics[width=0.32\linewidth]{figures/PPK_HI4PI/NHI_CNM_LOWER_HI4PI_mollview_\i.png}
  }
  \caption{Mollweide projection centered on the Galactic center of the lower limit on the cold gas mass fraction $f^{k_{v,\rm lim}}_{\rm low, HI4PI}$ (top) and the lower limit on the cold gas column density $f^{k_{v,\rm lim}}_{\rm low, HI4PI}\,\NHI^*$ in units of 10$^{19}$\,cm$^{-2}$ (bottom) of the entire HI4PI dataset in the velocity range $-90 < v < 90$\,\kms.\am{fixme}
}
  \label{fig:FMIN_CNM_LOWER_HI4PI_mollview}
\end{figure*}

In Figure~\ref{fig:FMIN_CNM_LOWER_HI4PI_mollview} we present the same two quantities but for nine $k_{v,\rm lim}$ values sampled in the range $0.01<k_v\,(\kms)^{-1} <0.27$.

\bibliography{ppv-fft}

\end{document}